\pdfoutput=1

\documentclass[fleqn,usenatbib]{mnras}

\newcommand{\n}[1]{}
\newcommand{\footnope}[1]{}

\RequirePackage{snapshot}
\usepackage{newtxtext,newtxmath}
\usepackage{placeins}

\usepackage[T1]{fontenc}
\usepackage{ae,aecompl}
\usepackage[dvipsnames]{xcolor}

\usepackage{graphicx}	%
\usepackage{amsmath}	%
\usepackage{amssymb}	%

\title[Resolved galaxy scaling relations in EAGLE]{Resolved galaxy scaling relations in the {\sc eagle} simulation: star formation, metallicity and stellar mass on kpc scales}

\author[J. W. Trayford]{
James W. Trayford$^{1}$\thanks{E-mail: trayford@strw.leidenuniv.nl} and Joop Schaye$^{1}$
\\
$^{1}$Leiden Observatory, Niels Bohrweg 2, 2333 CA Leiden, Netherlands\\
}

\date{Accepted XXX. Received YYY; in original form ZZZ}

\pubyear{2018}

\begin{document}
\defcitealias{Abdurrouf18}{A18}
\defcitealias{Wuyts13}{W13}
\defcitealias{Hsieh17}{H17}
\defcitealias{CanoDiaz16}{CD16}
\label{firstpage}
\pagerange{\pageref{firstpage}--\pageref{lastpage}}
\maketitle

\begin{abstract}
We explore scaling relations between the physical properties of spatially resolved regions within the galaxies that emerge in the \textit{Evolution and Assembly of GaLaxies and their Environments} (EAGLE) hydrodynamical, cosmological simulations. Using 1~kpc-scale \textit{spaxels}, we compute the relationships between the star formation rate and stellar mass surface densities, i.e. the spatially resolved star-forming main sequence (rSFMS), and between the gas metallicity and the stellar mass surface density, i.e. the spatially resolved mass-metallicity relation (rMZR). We compare to observed relations derived from integral field unit surveys and galaxy imaging. EAGLE reproduces the slope of the local ($z \approx 0.1$) rSFMS well, but with a $\approx-0.15$~dex offset, close to that found for the galaxy-integrated relation. The shape of the rMZR agrees reasonably well with observations, replicating the characteristic turnover at high surface density, which we show is due to AGN feedback. The residuals of the rSFMS and rMZR are negatively (positively) correlated at low (high) surface density. The rSFMS becomes shallower as the simulation evolves from $z=2$ to 0.1, a manifestation of \textit{inside-out} galaxy formation. The shape of the rMZR also exhibits dramatic evolution, from a convex profile at $z=2$ to the observed concave profile at $z=0.1$, such that the gas in regions of high stellar density is more enriched at higher redshift. The redshift independence of the relationship between the galaxy-wide gas fraction and metallicity in EAGLE galaxies is not preserved on 1~kpc scales, implying that chemical evolution is \textit{non-local} due to the transport of gas and metals within galaxies. %
\end{abstract}

\begin{keywords}
galaxies: formation - galaxies: evolution - galaxies: structure
\end{keywords}

\section{Introduction}
\label{sec:intro}
\vspace{1.5ex}

The mere existence of \textit{`scaling relations'} (i.e. trends between observables and/or inferred physical properties)  amongst widely separated galaxies suggests a commonality in their formation processes. Scaling relations encode information about key physical processes in galaxies, and provide touchstones for theoretical models of galaxy formation and evolution. 

One such scaling relation is the so-called \textit{star-forming main sequence} (SFMS, \citealt{Noeske07}); a relatively tight\footnote{With a spread of $\approx$0.2-0.35~dex.}, evolving relationship between the integrated stellar masses ($M_\star$) and star formation rates (SFR) of actively star-forming galaxies, observed from $z=0-6$ \citep[e.g.][]{Brinchmann04, Rodighiero11, Elbaz11, Speagle14}. This relation is typically well fit by a power law, ${\rm SFR} \propto M_\star^{n}$. The normalisation has been found to evolve dramatically, with typical SFRs a factor $\approx 20$ higher at $z=2$ relative to the present day \citep[e.g.][]{Whitaker12}. The evolution of the index (or `slope') is less clear, but typically values of $n=0.6-1$ are recovered \citep[for a compilation of observations, see e.g.][]{Speagle14}. 

The SFMS tells us about the ongoing growth of galaxies in the Universe, and its existence is suggestive of self regulation, where the inflow and outflow of gas are balanced by the influence of feedback processes \citep[e.g.][]{Schaye10, Bouche10, Dave11, Lilly13, Tachella16}. A possible flattening of the high-mass end has been explained as a physical effect due to the influence of quenching processes \citep[e.g.][]{Keres05, Dekel06, Croton08}, or simply as a manifestation of higher bulge-to-disc ratios in massive galaxies \citep[e.g.][]{Abramson14, Schreiber15}. However, \citet{Whitaker15} suggest the flattening may be spurious, due to a poor separation of the active and passive populations .

The relationship between $M_\star$ and gas-phase metallicity (MZR) complements the SFMS, and has also been widely studied using observations of local galaxies \citep[e.g.][]{Tremonti04, Kewley08} and those at high redshift \citep[e.g.][]{Erb06, Henry13, Maier14, Zahid14}. Metallicity is generally found to increase  with $M_\star$ at low masses, but to plateau or turn over for galaxies with $\log_{10}(M_\star /{\rm M_\odot}) \gtrsim 10.5$. Metals are considered to play a key physical role in the star formation process, as more enriched gas can is more efficient at cooling, and metals deposited in dust provide sites for molecule formation in the ISM. The ISM is enriched by previous stellar generations and can be ejected by galactic winds, and thus gas-phase metallicity ($Z_{ \rm gas}$) also encodes information about the star formation and outflow history of a galaxy. 

The multivariate relationship between $M_\star$, SFR and $Z_{\rm gas}$ provides insight into key aspects of galaxy formation. \citet{Mannucci10} and \citet{LaraLopez10} identify a \textit{`fundamental'} relationship between these properties, showing an anti-correlation between metallicity and the residuals of the $M_\star$-SFR relation. This relation has been found to exhibit little evolution \citep[e.g.][]{Stott13, Hunt16}, though some studies have observed this to break down at high redshift \citep[$z \gtrsim 2$,][]{Mannucci10,Salim15}. The strong dependence of metallicity on \textit{instantaneous} SFR (as opposed to the total integrated star formation) lends support to the model of galaxies as existing in a self-regulated dynamic equilibrium of inflowing and outflowing gas \citep[e.g.][]{Finlator08, Dave12, DeRossi17, Torrey17}.

While scaling relations such as the SFMS and MZR are typically devised in terms of integrated galaxy properties, they are regulated by the star formation and feedback processes taking place on sub-galactic scales. As such, spatially resolving the properties of \textit{regions} of galaxies may provide further insight into how the integrated relations arise. \citet{Wuyts13} use resolved HST imaging to find a relation between the surface density of stars ($\Sigma_\star$) and star formation ($\Sigma_{\rm SFR}$) in $0.7 \leq z < 1.5$ galaxies analogous to the SFMS, suggesting that the relationship between stellar mass and star formation rate holds down to 1~kpc scales: a resolved star-forming main sequence (rSFMS). %
However, the profiles of galaxies are found to be inherently clumpy for these scales and redshifts \citep{ForsterSchreiber11, Genzel11, Wisnioski15}, so it was not obvious that such a relation would hold for the Hubble-type galaxies dominant locally. 

Integral Field Unit (IFU) instruments can be used to measure resolved scaling relations in local galaxies.  IFU surveys have now yielded spatially resolved spectroscopy for considerable galaxy samples in the local Universe \citep[e.g.][]{Sanchez12,Bryant15,Bundy15}. IFUs can be used to spatially map star formation via optical proxies such as H$\alpha$ luminosity \citep[e.g.][]{Kennicutt98}, gas phase metallicity through emission line ratios \citep[e.g.][]{Sanchez17}, and stellar mass through SED fitting techniques \citep{Sanchez16, Goddard17}. %
Multiple IFU studies of local ($z < 0.1$) galaxies analysing the rSFMS relation \citep{CanoDiaz16,Hsieh17,Medling18} and the resolved MZR  \citep[rMZR, e.g.][]{RosalesOrtega12, BarreraBallesteros17} are now available.

While integrated scaling relations have been widely used to test and calibrate cosmological galaxy formation models with statistically significant galaxy populations, resolved scaling relations have not yet been used for these purposes.  Resolved relations could be particularly constraining for large-volume hydrodynamical simulations, as galaxy properties arise from the local hydrodynamical calculation\footnote{As opposed to  semi-analytic models where galaxy profiles are typically imposed.}, and subgrid models for unresolved physics. It is possible for a simulated galaxy population to reproduce integrated relations, while failing to yield galaxies with realistic internal structures \citep[e.g.][]{Crain15}. Spatially resolved scaling relations are thus useful diagnostics of both the structure and demographics of simulated galaxy populations, complementing population comparison studies of galaxy morphology \citep[e.g.][]{Croft09, Sales10, Snyder15, Lagos17a, Dickinson18, Trayford18}, property gradients \citep[e.g.][]{Cook16, Taylor17, Tissera18} and sizes \citep[e.g.][]{Mccarthy12, Bottrell17, Furlong17, Genel18}.

If simulations can broadly match both integrated and resolved observations simultaneously, then they may provide insight into their physical origin. Simulations afford the opportunity to follow virtual galaxies through cosmic time and assess how star formation, feedback, environmental effects and angular momentum evolution build their structural properties. While these relationships are likely complex in detail, they may yield effects that are conceptually simple. For example, simulations can show whether `inside-out' star formation is consistent with the evolution of the resolved main sequence, as is hinted at by empirical models \citep[e.g.][]{Abdurrouf18, Ellison18}.

In this study, we explore spatially resolved scaling relations using the EAGLE simulation suite  \citep{Schaye15,Crain15,McAlpine16}. By mapping gas and stellar properties of simulated galaxies, we assess how well the simulation reproduces the local relations and how they become established. We focus in particular on the rSFMS  and rMZR. In section~\ref{sec:eagle} we briefly describe the aspects of the simulation most relevant for this study. Section~\ref{sec:obs} then describes how we construct resolved property maps and attempt to emulate the selection effects of contemporary IFU studies. We present results in section~\ref{sec:res}, comparing the relations resolved on 1~kpc scales to observations at $z=0.1$, and showing how galaxies of different mass contribute to the overall relation. Section~\ref{sec:evo} then focuses on evolution, showing how the kpc-scale relations vary with redshift, as well as how this relates to the integrated relation and galaxy profiles. Finally, we summarise our conclusions in section~\ref{sec:summary}. Unless stated otherwise, all distances are in proper (as opposed to comoving) coordinates.

\section{Simulations}
\label{sec:eagle}

In this study we utilise the \textit{Evolution and Assembly of GaLaxies and their Environments} (EAGLE) suite of cosmological, hydrodynamical simulations \citep[][]{Schaye15,Crain15}. We focus on two volumes in particular: the fiducial 100$^3$~Mpc$^3$ box (Ref100) and the higher-resolution recalibrated 25$^3$~Mpc$^3$ box (RecHi25). The Ref100 (RecHi25) simulation resolves baryonic matter with an initial particle mass of of $m_g = 1.81\times10^6  {\rm M_\odot}$ ($m_g = 2.25\times10^5 {\rm M_\odot}$). The RecHi25 simulation has a factor 2 higher spatial resolution than Ref100 (see Table~\ref{tab:specs}). A \textit{Planck-1} cosmology is assumed by the simulations and throughout this work \citep{Planck}.

EAGLE follows the co-evolution of baryons and dark matter in each volume using a modified version of the {\sc Gadget-3} TreeSPH code (an update to {\sc Gadget-2}, \citealt{Springel05b}), with star formation and feedback models parametrised differently. Changes to the hydrodynamics calculation include a pressure-entropy formulation \citep{Hopkins13}, artificial viscosity and conduction switches \citep[][]{Cullen10, Price08}, a \citet{Wendland95} C$_2$ smoothing kernel, and the timestep limiter of \citet{Durier12}. These enhancements are described in \citet{Schaye15} and \citet{Schaller15}.

Additional physics models are included for a number of key processes. Star formation is sampled stochastically in gas above a metallicity-dependant density threshold \citep{Schaye04}, using a pressure dependent adaptation of the observed Kennicutt-Schmidt law \citep{Schaye08} to compute star formation rates. Gas particles are converted wholesale into simple stellar populations, inheriting the initial mass and chemical composition of their progenitor gas.

Stellar mass loss and enrichment follow the implementation of \citet{Wiersma09b} for each star particle, distributing ejected material over the SPH neighbours. Nine elements are tracked explicitly: H, He, C, N, O, Ne, Mg, Si and Fe. Nucleosynthetic yields from winds, core-collapse supernovae and SNIa follow \citet{Portinari98}, \citet{Marigo01} and \citet{Thielemann03}. Photoheating and cooling rates are computed for each of these elements individually \citep{Wiersma09a}. Thermal feedback associated with both star formation \citep{DallaVecchia12} and AGN is implemented stochastically. Feedback parameters were calibrated to reproduce the $z=0.1$ galaxy stellar mass function, mass-size relation and black hole mass stellar mass relation.  

Dark matter halos are identified using a Friends-of-Friends (FoF) algorithm, and their constituent self-bound substructures (subhalos) are identified using the SUBFIND code \citep{Springel01,Dolag09}. For this work, galaxies are taken to be exclusive to individual subhalos, and within a 30~proper~kpc (pkpc) spherical aperture about the galaxy centre to mimic a Petrosian aperture \citep{Furlong15}. Our galaxy centring method uses a shrinking spheres approach, following \citet{Trayford18}.

\section{Comparing to observations}
\label{sec:obs}

\begin{table}
\caption{Specifications of the two primary EAGLE simulation and the three contemporary IFU surveys considered in this work. EAGLE provides two discrete snapshots at comparable redshifts. As these are predominately the same galaxies $\approx 1$~Gyr apart, we quote only the number of coeval galaxies at $z=0$. The chosen surveys all resolve physical scales of $\approx$1~kpc, well matched to the resolution of EAGLE.}
\label{tab:specs}
\begin{center}
\begin{tabular}{lccc}
\hline
 & Min. scale$^{\rm a}$ & Redshifts & Selection$^{\rm b}$ \\
 & (kpc) & ($z$) & (criterion $\vert$ number)\\
\hline\hline
Ref100 & $\approx 0.7$ & $[0,0.1]$& $M_\star > 10^{9}{\rm M_\odot}$ $\vert$ $10^{4.1}$\\ 
RecHi25 & $\approx 0.35$ & $[0,0.1]$& $M_\star > 10^{8.1}{\rm M_\odot}$ $\vert$ 620\\ 
\hline
CALIFA & $0.8-1.0$ & $0.005-0.03$ & $45^{\prime\prime} < {D_{25}}^{\rm c}<80^{\prime\prime}$ $\vert$ 600\\ 
MaNGA & $1.3-4.5$ & $0.01-0.15$ &  $M_\star > 10^{9}{\rm M_\odot}$ $\vert$ $10^{4}$\\
SAMI & $1.1-2.3$ & $0.004-0.095$ &  $M_\star > 10^{8.2}{\rm M_\odot}$ $\vert$ 3400\\
\hline
\end{tabular}
\end{center}
{\footnotesize $^{\rm a}$  For EAGLE, this is the Plummer-equivalent maximum gravitational softening.\\
$^{\rm b}$ %
Observed selection functions are not complete in mass. More detail on galaxy selection can be found in \S~\ref{sec:select}.\\
$^{\rm c}$ $D_{25}$ is the $r$-band 25 mag arcsec$^{-2}$ isophotal diameter.
}

\end{table}

In this study we focus on comparing with the results from two contemporary IFU surveys in particular: CALIFA \citep{Sanchez12} and MaNGA \citep{Bundy15}. Results from these campaigns are particularly well-suited for comparison with EAGLE\footnote{Other IFU surveys, such as SAMI \citep{Bryant15}, are similarly compatible. While we compare with certain published resolved scaling relations in this study, future comparison with such surveys would be informative.}, as detailed below. Some specifications of the surveys and simulated data are listed in Table \ref{tab:specs}. 
 
These surveys sample galaxies on spatial scales of $\approx $1~kpc. This scale is well-matched to the standard resolution limit in EAGLE, where structure formation is suppressed on scales $\lesssim 0.7$~kpc due to gravitational smoothing. 

MaNGA provides a sample of $\sim 10^4$ galaxies with masses $M_\star > 10^9 {\rm M_\odot}$, comparable to the $\approx 13,000$ galaxies above this mass limit in the Ref100 simulation volume at both $z = 0$ and $z=0.1$. CALIFA offers a smaller sample, selecting more local sets of galaxies using a $M_\star > 10^{8.2} {\rm M_\odot}$ mass cut and apparent size selection respectively. As $10^{8.2} {\rm M_\odot}$ is equivalent to $\sim 100$ star particles at standard EAGLE resolution, such galaxies are likely insufficiently resolved in the simulation. Resolution effects and convergence are explored directly in Appendix~\ref{sec:convergence}. We discuss emulating galaxy and \textit{`spaxel'} selection effects in section~\ref{sec:select}. 

\subsection{Property maps}
\label{sec:maps}
This study requires spatially resolved maps of physical properties of EAGLE galaxies. We employ the publicly available {\tt py-sphviewer} code \citep{BenitezLambay15}, which uses adaptive kernel smoothing to create smooth two-dimensional imaging from sets of discrete three-dimensional particles. Galaxies are mapped individually, extracting material for a given subhalo and applying a 30~pkpc spherical aperture about the galaxy center. The maps are made at an intrinsic $256\times256$ \textit{`spaxel'} resolution for a $60\times60$~kpc field and in three projections: simulation $xy$-coordinates, face-on and edge-on. The face-on and edge-on orientations are defined via the primary baryonic rotation axis. Galaxy centering and orientation follow the procedures detailed in \citet{Trayford17}.

When considering spatially resolved properties, the manner in which the particles are smoothed may influence our results. In the adaptive smoothing case, a smoothing length is applied to each particle. For the gas, the obvious choice of smoothing length is the SPH kernel  used in the simulation's hydrodynamical calculations. Stars, however, are not subject to hydrodynamic forces. As no physical scale is precomputed for adaptive smoothing between stars, smoothing lengths are calculated for each star particle to enclose a fixed number of neighbours. The choice of this number is a compromise between mitigating both granularity in the stellar profiles and washing out spatial trends.  We compute stellar smoothing lengths based on the 64th nearest neighbour, matching the radiative transfer imaging described in \citet{Trayford17}.

In order to test the influence of smoothing, we also make sets of images where the smoothing of stars and gas are set to zero, such that a particle contributes solely to the spatial bin in which it resides. By comparing results with and without smoothing, we can test whether the choice of smoothing scale is important. Suffice to say, the influence of intrinsic smoothing of the gas and stellar material on our results is small, but explored in more detail in Appendix~\ref{sec:pointlike}.

The property maps made for the stars and star-forming gas in each galaxy are listed in Table~\ref{tab:prop}. These properties are either weighted or mapped directly (in the case of masses themselves and star formation rates), and are stored for all spaxels with non-zero mass. For this study, all maps are re-binned onto a factor 4 coarser grid such that the spaxels sample kpc scales (0.9375~kpc). This mapping scheme is applied to a selection of EAGLE galaxies, as detailed below.

\begin{table}
\begin{center}
\caption{Summary of the property maps produced for EAGLE galaxies. Maps are produced for each galaxy in three orientations with 256x256 \textit{`spaxels'} over a 60x60~kpc field of view. Unless stated otherwise, the maps are re-binned to a spatial resolution of $\approx$1~kpc (0.9375~kpc) for comparison to data. The properties used in this work are stellar mass, SFR and the star-forming (SF) gas-phase elemental abundances of O and H. The last column indicates how properties are aggregated within a spaxel, with \textit{`weighted average'} indicating the stellar mass and SFR weighted mean values respectively for stars and gas.}
\label{tab:prop}
\begin{tabular}{lccc}
\hline
 & Stars & SF gas & Spaxel value\\
\hline
Mass & \checkmark &  \checkmark &  Sum \\
SFR &  - &  \checkmark &  Sum \\
Density & - &  \checkmark & Mean\\
Metallicity & \checkmark  & \checkmark & Weighted average\\
Abundances$^{\rm a}$ & \checkmark & \checkmark & Weighted average\\
LoS velocity & \checkmark & \checkmark & Weighted average\\
Age & \checkmark & - & Weighted average\\
\end{tabular}
\end{center}
{\footnotesize $^{\rm a}$ Mapped elements are H, C, N, O and Fe }
\end{table}

\subsection{Selection effects}
\label{sec:select}
Selection effects are a key consideration for a robust comparison to data. For spatially resolved surveys, these effects are induced by the selection of both the galaxies  and the spaxels that sample them. We detail our attempts to imitate observational selection effects below.

\subsubsection{Galaxy selection}
\label{sec:galsel}

The target selection modelling described in this section is used for comparison to local IFU surveys in \S\ref{sec:res}. Galaxy selection of IFU surveys is generally more complex than that of imaging surveys, which are typically complete down to some stellar mass or flux limit.

While MaNGA employs a lower mass limit, the $M_\star$ selection functions of both surveys differ from a simple mass cut. In MaNGA the galaxy selection is designed to be uniform in $\log_{10}(M_\star)$ for $\log_{10}(M_\star/M_\odot) > 9$, in order to sample galaxies across a range in stellar mass \citep{Bundy15}. The MaNGA distribution stays approximately uniform in $\log_{10}(M_\star)$ up to  $\log_{10}(M_\star/M_\odot) \approx 11.3$, above which the number density drops rapidly \citep{Wake17}. A uniform selection reduces the number of spaxels contributed by lower-mass objects and boosts that of higher-mass objects relative to a $M_\star$-complete survey. 

An intuitive  way to reproduce a uniform selection with EAGLE would be to select galaxies with a probability inversely proportional to the galaxy stellar mass function, i.e. $P(M_\star) \propto \phi(M_\star)^{-1}$.  However, the low number counts of high-$M_\star$ galaxies in EAGLE, an inherently volume-limited sample, means that a very small fraction of galaxies would tend to be selected. 

To enable better utilisation of the available EAGLE galaxy population, we instead emulate a flat galaxy selection in the $z=0.1$ Ref100 population using a hybrid method; we stochastically select galaxies to be uniform in $\log_{10}(M_\star)$ between $9 \leq \log_{10}(M_\star/{\rm M_\odot}) < 10$, while all galaxies of $\log_{10}(M_\star/{\rm M_\odot}) \geq 10$ are selected,  and their spaxel contributions are weighted appropriately to mimic a uniform selection at high $\log_{10}(M_\star)$. %
Together, this provides a sample of 9284 galaxies, comparable to that of MaNGA. %
The fiducial weighting scheme, $w$, applied for each galaxy is then 
\begin{equation}
\label{eq:w100}
  w_{\rm 100}(M_\star) =
  \begin{cases}
    \frac{\phi_{100}(10^{10} \, {\rm M_\odot})}{\phi_{100}(M_\star)} & \text{if $10^{10} \leq M_\star/{\rm M_\odot} < 10^{11.3}$}\\
    \frac{\phi_{100}(10^{10}\, {\rm M_\odot})}{\phi_{100}(10^{11.3}\, {\rm M_\odot})} & \text{if $M_\star/{\rm M_\odot} \geq 10^{11.3}$},    
  \end{cases}
\end{equation}
where $\phi_{100}(M_\star)$ is the Ref100 galaxy stellar mass function, binned in $\log_{10}(M_\star/{\rm M_\odot})$. For the 35 galaxies above the upper mass limit of $\log_{10}(M_\star/{\rm M_\odot}) = 11.3$ the weighting value saturates at $w_{100} \approx 25$, and the galaxy contribution falls away with the galaxy stellar mass function. For a 25~Mpc simulation box like RecHi25, the volume is 64 times smaller, so all galaxies with $\log_{10}(M_\star/{\rm M_\odot}) \geq 9$ are selected (261 systems in RecHi25), with weighting %
\begin{equation}
\label{eq:w25}
  w_{\rm 25}(M_\star) = \frac{\phi_{25}(10^{10})}{\phi_{25}(M_\star)},
\end{equation}
where $\phi_{25}(M_\star)$ is the RecHi25 galaxy stellar mass function.

As no galaxies are found with $\log_{10}(M_\star/{\rm M_\odot}) > 11.3$ in the 25~Mpc volumes, no saturation value for $w_{25}$ is enforced. Throughout, the Ref100 and RecHi25 samples are treated separately.   

In addition to this fiducial weighting scheme, we alternatively weight the same sample of EAGLE galaxies to represent the CALIFA mass distribution \citep[e.g.][]{GarciaBenito15}. This mass distribution is peaked at $\log_{10}(M_\star/{\rm M_\odot}) \approx 10.8$, such that pixels from these galaxies contribute more weight than higher- and lower-$M_\star$ galaxies, relative to the MaNGA weighting scheme. We show the relations arising from this alternative weighting scheme in a number of the plots of \S\ref{sec:res} and Appendix~\ref{sec:pfrac}. We note that we do not replicate the projected size selection used to produce the CALIFA galaxy sample. 

Alongside the $M_\star$ target selection, an effective cut in specific SFR is also typically employed by IFU studies \citep[e.g.][]{CanoDiaz16, Hsieh17}. This allows the same galaxies to contribute to the rSFMS as contribute to the integrated SFMS, but also has the practical motivation of isolating galaxies with primary ionisation mechanisms attributable to star formation (rather than shocks or AGN). Given the known -0.2~dex offset of the EAGLE integrated SFMS from typical observational studies \citep{Furlong15}, employing the stringent observed cut bisects the integrated SFMS. We instead use the $\log_{10}({\rm sSFR/yr^{-1}}) > -11$ cut of \citet{Furlong15} to isolate \textit{`star-forming'} EAGLE galaxies\footnote{We note that the sSFR criterion may differ for rMZR studies \citep[e.g.][]{Sanchez17}, but as we find the influence is minimal on the EAGLE rMZR, we use the same cut for consistency.}.

\subsubsection{Spaxel selection}
\label{sec:spaxsel}

With the galaxy selection and weighting scheme in place, we now consider which spaxels contribute to the scaling relations. For the rSFMS and rMZR studied here, we require spaxels to sample both star-forming gas and stellar populations. To roughly mimic the observed relations, we employ $\dot{\Sigma}_\star$ and $\Sigma_\star$ limits.

Observationally, star formation rates are typically inferred from (dust-corrected) H$\alpha$ luminosities, $L_{\rm H\alpha}$, via a scaling of
\begin{equation}
\dot{M}_\star = 7.49\times 10^{42} \; {\rm M_\odot} \; {\rm yr^{-1}}f_{\rm IMF} \frac{L_{\rm H\alpha}}{\rm erg \; s^{-1}} ,
\end{equation}
where $f_{\rm IMF}$ is a factor accounting for differences between IMF assumptions. For the \citet{Chabrier03} IMF assumed by EAGLE, we use $f_{\rm IMF} = 1.57$ \citep[e.g.][]{Lacey16}. 

Star formation is discernible in CALIFA for spaxels with star formation rate surface densities of $\dot{\Sigma}_\star  \gtrsim 10^{-9.} \; {\rm M_\odot \; yr^{-1} \; pc^{-2}} $ \citep[Fig. 2 of][]{CanoDiaz16}. In MaNGA, star formation rate surface densities of $\dot{\Sigma}_\star \gtrsim 10^{-10} \; {\rm M_\odot \; yr^{-1} \; pc^{-2}}$ are detected \citep{Hsieh17}. To represent this selection effect, we employ the MaNGA-like criterion of  $\dot{\Sigma}_\star  > 10^{-10} \; {\rm M_\odot \; yr^{-1} \; pc^{-2}}$ when selecting spaxels that contribute to the EAGLE relations.

The independent variable in the scaling relations considered here is the stellar mass surface density, $\Sigma_\star$, so we can simply compare EAGLE to observations over a range where this is reliable for both samples.  We compare plots in the range $\Sigma_\star > \, 10\ {\rm M_\odot \; pc^{-2}}$, equivalent to a kpc-scale spaxel sampling about 5 star particles at Ref100 resolution\footnote{Typically, more than 5 particles contribute to those spaxels, due to the particle smoothing.}. While the IFU surveys all probe down to lower $\Sigma_\star$ values, this limit mitigates stochastic effects due to poor particle sampling in EAGLE. 

Another important factor is the radial coverage of spaxels. IFU instruments have limited angular size, and sample the inner regions of galaxies. Typically, the upper radial limit out to which gradients are measurable is $\lesssim 3$ times the effective radius \citep{Bundy15}. Observationally the effective radius, $R_e$, represents the half-light radius, but here we take $R_e$ to be the projected half-mass radius \citep[see][]{Furlong17}, selecting only spaxels at radii $< 3 R_e$. In practice, the radial cut has no perceptible effect on our result for the stellar surface density regime of $\Sigma_\star > 10\ {\rm M_\odot \; pc^{-2}}$ considered in our plots. %

\subsection{Metallicity calibration}
\label{sec:zcal}

Estimating metallicities observationally is highly challenging, and metallicity calibrations are subject to considerable uncertainty. For our study, we separate two broad classes of systematic uncertainty. One is absolute calibration, i.e. how well the overall abundance of heavy elements in stars and gas can be inferred. Another pertains to the relative calibration, i.e. how well observable metallicity indicators trace the underlying metallicity variation between, or within, galaxies. While a detailed discussion of metallicity calibration is beyond the scope of this study \citep[see instead e.g.][]{Kewley08}, we discuss some of the most pertinent aspects here.

The EAGLE simulations use the nucleosynthetic yields of \citet{Portinari98} and \citet{Marigo01} for stellar evolution and core-collapse supernovae, as well as the SNIa yields of \citet{Thielemann03} \citep[with some modification, see][]{Wiersma09b}, which dictate the absolute abundances of chemical elements in gas and stars. As discussed in \citet{Wiersma09b}, even for a fixed IMF, the yields are uncertain at a 0.3~dex level. The simulations yield good agreement with integrated mass-metallicity relations for stars and gas \citep{Tremonti04,Zahid14, Gallazzi09} for high stellar masses ($M_\star > 10^{10} {\rm M_\odot}$ for Ref-100 and $M_\star > 10^{9} {\rm M_\odot}$ for Recal-25, see \citealt{Schaye15}). Assuming $12 + \log_{10}({\rm O/H})_\odot = 8.69$ \citep{Allende01}, typical metallicities become super-solar for $\log_{10}(M_\star/{\rm M_\odot}) \gtrsim 9.5$, saturating at around three times the solar value. 

\citet{Sanchez16} present a spatially resolved gas-phase mass-metallicity relation and explore a number of metallicity calibrations using CALIFA, with the same pipeline also used to derive the relation in MaNGA data \citep{BarreraBallesteros17}. Fig.~3 of \citet{Sanchez16} demonstrates the integrated mass-metallicity relation for a multitude of calibrators, but finds values systematically lower than that of both EAGLE and other observational studies by $> 0.2$~dex for all calibrators. Part of this discrepancy may be attributable to anchoring the metallicity values to empirical ionisation parameter measurements as opposed to photoionisation models  \citep{Sanchez16}.   

In order to compare to IFU data, we therefore apply a constant shift to the EAGLE $\log_{10}({\rm O/H})$ values of $-0.6$~dex, which represents the difference between the high-$M_\star$ plateau for the $R_{23}$ calibration of \citet{Sanchez16} and that of \citet{Zahid14}. The motivation for this recalibration is that it brings the absolute calibration of the EAGLE data into agreement with IFU studies for the same metallicity indicator. However, we follow recent theoretical work on metallicity evolution in galaxies and put no emphasis on absolute metallicities, given the large uncertainties \citep[e.g.][]{Torrey17}. Instead, we only compare the shape of the relation, and evolution in metallicity in a relative sense. \citet{Sanchez16} show that despite the shift in absolute calibration, the majority of indicators yield integrated mass-metallicity relations with very similar shapes. This gives us some confidence in the validity of the  relative calibration for a comparison between the shape of the resolved mass-metallicity relation in EAGLE and the data.

\section{scaling relations at low redshift}
\label{sec:res}

\begin{figure}
\includegraphics[width=\columnwidth]{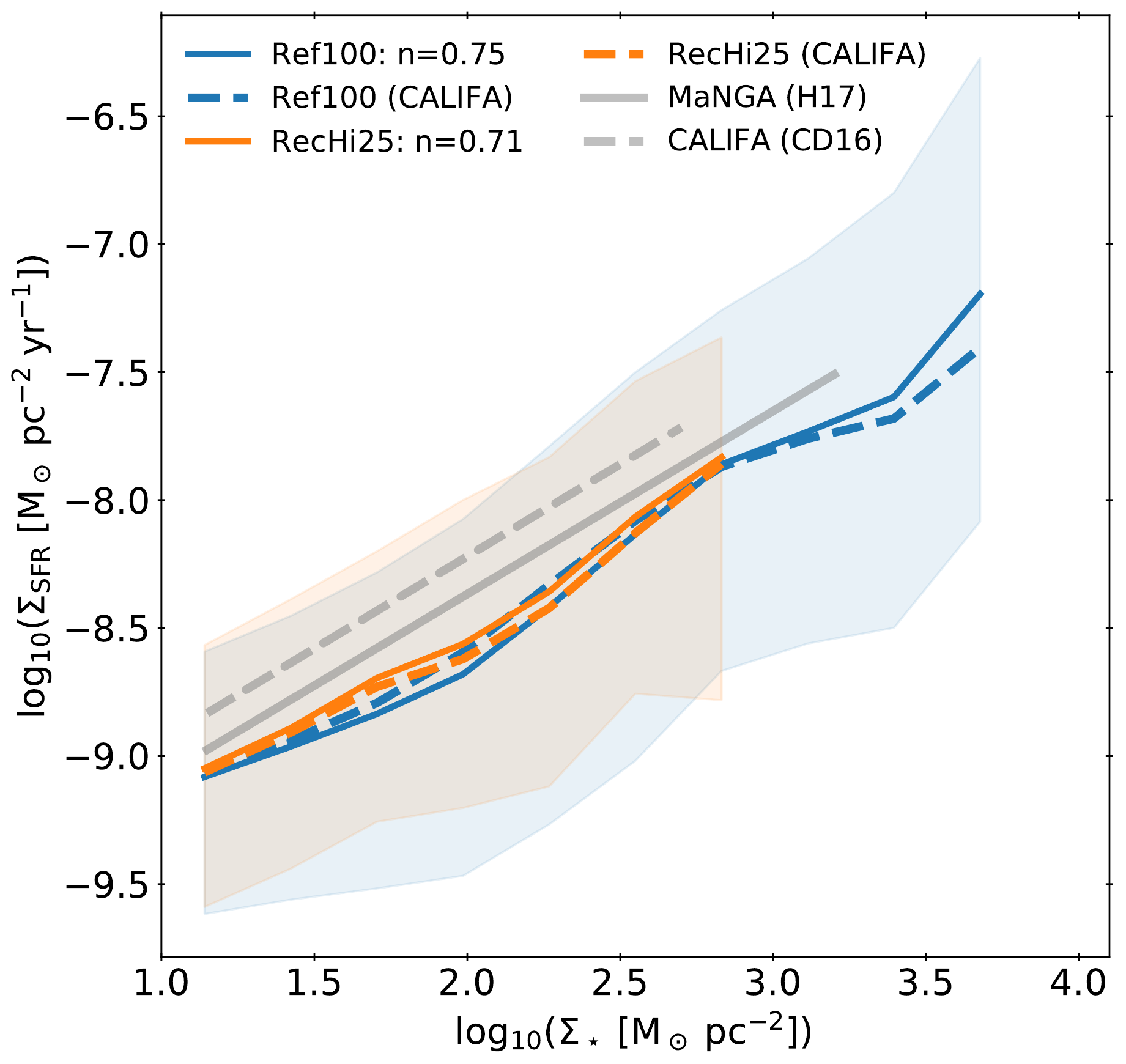}
    \caption{The spatially resolved star-forming main sequence relation (rSFMS) at $z=0.1$, 
    plotted using spaxels sampling 1~kpc scales. We compare to the observed relations 
    of \citet{CanoDiaz16} (CALIFA) and \citet{Hsieh17} (MaNGA), which sample approximately the same spatial scales at $z<0.1$. The observed rSFMS fits are 
    plotted for $\Sigma_\star/({\rm {M_\odot \; pc^{-2}}}) > 10^{1.15}$ (the centre of the lowest $\Sigma_\star$ bin), extending to the edge of the contour enclosing $80\%$ of the contributing spaxels in the $\Sigma_\star$-$\Sigma_{\rm SFR}$ plane \citepalias[see][]{CanoDiaz16, Hsieh17}. The weighted 
    median values are plotted for each  bin in which no single galaxy contributes $> 5 \%$ of the 
    total weighting. %
    Shaded regions enclose the 16th-84th percentiles of the 
    weighted spaxel distribution in each $\log_{10}(\Sigma_\star)$ bin.  We find that EAGLE reproduces the observed rSFMS slope well, with a $\approx-0.2$~dex offset in normalisation.}
    \label{fig:sfms}
\end{figure}

Having developed a procedure to measure resolved properties of EAGLE galaxies in a manner suitable for a first order comparison with low-redshift IFU surveys, we now present our results and compare directly to the observationally inferred relations. We first consider the $z=0.1$ rSFMS and rMZR in sections \ref{sec:orsr} and \ref{sec:mzr} respectively.  We then explore how galaxies of different mass ranges contribute to them in section \ref{sec:mdep}. Finally, we investigate the relationship between the rMZR and rSFMS by comparing their residuals in section \ref{sec:residuals}. Where appropriate, the observed $\Sigma_\star$ and $\Sigma_{\rm SFR}$ are corrected for consistency with the \citet{Chabrier03} IMF assumed by EAGLE. Convergence properties of the relations are investigated further in Appendix~\ref{sec:convergence}.

\subsection{The resolved star forming main sequence}
\label{sec:orsr}

In Fig.~\ref{fig:sfms} we plot the resolved star-forming main sequence (rSFMS) for the Ref-100 and Recal-25 simulations, colour-coded blue and orange respectively. Here, solid lines represent the \textit{`weighted median'} relations. These are calculated by finding the 50th percentile of the weighted (via Equations~\ref{eq:w100} and \ref{eq:w25}) distribution of $\Sigma_{\rm SFR}$ for galaxies in uniform, contiguous bins of $\log_{10}(\Sigma_\star)$. The default weighting scheme is intended to replicate a galaxy selection uniform in $\log_{10}(M_\star)$, approximating that of the MaNGA survey \citep[e.g.][]{Wake17}. We only plot bins to which $\geq 10$ galaxies contribute, and where each individual spaxel contributes $\leq 5\%$ of the total weight. The 16th-84th percentile range of this distribution is indicated by the shaded region. The dashed coloured lines denote an alternative spaxel weighting, intended to represent the CALIFA mass distribution sample (see \S\ref{sec:spaxsel}). We compare to the relations derived for MaNGA \citepalias{Hsieh17} and CALIFA \citepalias{CanoDiaz16}.%

Generally, we see that the fiducial (MaNGA-like) Ref100 rSFMS measured for EAGLE follows the observed slope well across the $1 \lessapprox \log_{10}\Sigma_\star/ ({\rm M_\odot \; pc^{-2}}) \lessapprox 3.2$ range, but with a normalisation $\approx0.15$~dex below the MaNGA relation \citepalias{Hsieh17}. This offset is consistent with the finding that the integrated SFRs of EAGLE galaxies display a $-0.2$~dex offset relative to the majority of observational studies \citep{Furlong15}. It is worth noting that there are observational studies which claim a lower normalisation of the integrated SFMS via SED fitting techniques \citep[e.g.][]{Chang15}, to which EAGLE agrees better, though there is still considerable uncertainty in the absolute normalisation of star formation rates.

Applying the alternative (CALIFA-like) weighting scheme to EAGLE (dashed lines) yields only marginal differences in the shape and normalisation of the relations. Comparing the observations, the CALIFA relation \citep{CanoDiaz16} is normalised $\approx 0.15$~dex higher than MaNGA over the plotted range, but is found to be consistent given the uncertainties \citep{Hsieh17}. The $\approx 0.3$~dex total offset of the Ref100 rSFMS below the CALIFA relation is thus compatible with a $\approx0.15$~dex offset as found for the EAGLE-MaNGA comparison, along with systematic uncertainties in the data. The consistency in the EAGLE relation recovered for both weighting schemes is somewhat reassuring. However, this does not necessarily indicate that the rSFMS is independent of $M_\star$, as is explored further in  \S\ref{sec:mdep}.  

The RecHi25 relation exhibits marginally closer agreement with the observations for the fiducial weighting scheme, reproducing the observed slope and exhibiting a $\approx0.15$~dex offset below the relation derived for MaNGA \citepalias{Hsieh17} over the probed $\Sigma_\star$ range. The fiducial RecHi25 relation is slightly higher than that of Ref100 (by $\lessapprox0.1$~dex), echoing the higher main sequence normalisation found for RecHi25 \citep[e.g.][]{Schaye15}. However, this difference is erased when a CALIFA-like weighting scheme is used. This may reflect that the CALIFA mass distribution peaks at $\log_{10}(M_\star/{\rm M_\odot}) \sim 10$, where differences between the Ref100 and RecHi25 sSFRs are minimal \citep{Furlong15}. The origin of these differences is explored further below; the influence of massive galaxies, which are not captured within the 25$^3$~Mpc$^3$ volume\ of RecHi25, are shown in \S\ref{sec:mdep}, and the minor influence of resolution is demonstrated in Appendix~\ref{sec:convergence}.

A power law ($\Sigma_{\rm SFR} \propto \Sigma_{\rm \star}^n$) is fit to the plotted Ref100 and RecHi25 weighted median values, and their slopes, $n$, are noted in the legend. We find that Ref100 (RecHi25) exhibits a slope of $\approx 0.71$ (0.75), close to the value of $\approx 0.72$ derived for MaNGA and CALIFA \citepalias{CanoDiaz16,Hsieh17}. The rSFMS slope has been found to be comparable to the integrated SFMS slope observationally. The literature compilation of \citet{Speagle14} predicts a SFMS slope that is slightly shallower than the rSFMS, with a value of 0.53 at $z=0.1$, though some individual studies give slopes varying from significantly steeper to shallower values. Considering active galaxies (${\rm SFR}/M^\star > 10^{-11} \, {\rm yr}^{-1}$) in the Ref100 run, we find a SFMS slope that is slightly shallower than the rSFMS, with a value of 0.6 for the mass range $9 < \log_{10}(M_\star/{\rm M_\odot}) \leq 11.3$, assuming the MaNGA-like galaxy weighting of equation~\ref{eq:w100}. 

The relationship between the integrated and resolved star-forming main sequence is complex, because it builds in the mass dependence of sizes, ISM distributions and star formation efficiency. Hence, interpreting the difference between the slopes of the SFMS and rSFMS is difficult. However, the different slopes of the rSFMS constructed from galaxies at different epochs, or from galaxies with differing levels of SFR for their stellar mass, have been used as an indicator of the \textit{inside-out} evolution of galaxies \citep[e.g.][]{Abdurrouf17, Medling18, Liu18, Ellison18}. Such evolution is explored  further in \S\ref{sec:evo}.

\subsection{The resolved mass-metallicity relation}
\label{sec:mzr}

\begin{figure}	\includegraphics[width=\columnwidth]{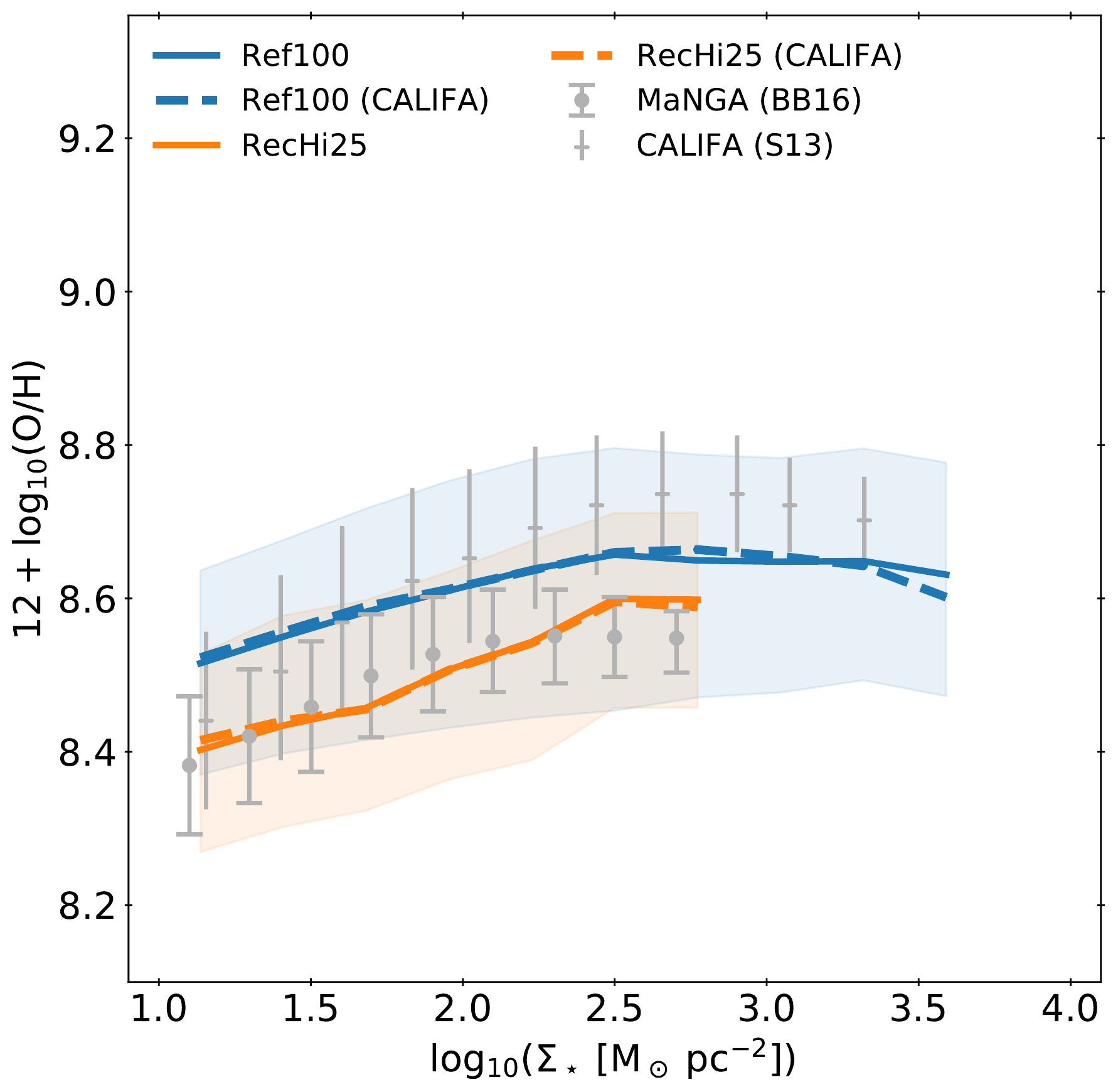}
    \caption{The weighted median spatially resolved relation between gas-phase metallicity and stellar mass (rMZR) for $z=0.1$ EAGLE galaxies. Because of uncertainties in absolute metallicity calibration, we consider only the shape of the relation and the EAGLE data has been shifted by -0.6~dex to account for calibration of absolute abundances, see \S\ref{sec:zcal} for details. \textit{Lines} represent  EAGLE, with colours and line-styles indicating simulation volume and spaxel weighting scheme respectively, as in Fig.~\ref{fig:sfms}. We compare to the local IFU studies of \citet{BarreraBallesteros16} (MaNGA) and \citet{Sanchez13} (CALIFA), grey circles and dashes respectively, where bars indicate the inferred 1$\sigma$ scatter.  We see that the predicted relations are qualitatively similar to the observations.}
    \label{fig:mz}
\end{figure}

We now consider the resolved relation between gas-phase metallicity and stellar mass (rMZR) for the Ref-100 and Recal-25 EAGLE simulations, plotted in Fig.~\ref{fig:mz}. This is constructed using the $\Sigma_\star$ and \textit{SF-weighted} gas-phase oxygen abundance, $\log_{10}({\rm O/H})$, of individual 1~kpc scale virtual spaxels. As for Fig.~\ref{fig:sfms}, the galaxy and spaxel selection are as described in sections~\ref{sec:galsel} and \ref{sec:spaxsel}, respectively. We do not concern ourselves with how well EAGLE captures the absolute calibration (i.e. normalisation) of this relation, given the uncertainties discussed in \S\ref{sec:zcal}. For ease of comparison, we thus applied a shift of -0.6~dex to the EAGLE results (see \S\ref{sec:zcal} for details).

Comparing first the Ref100 simulation to observations, we find that the shapes of the relations are similar. For $\Sigma_\star \lesssim 10^{2} {\rm M_\odot \; pc^{-2}}$, we find a positive trend where $\log_{10}({\rm O/H})$ increases by $\approx 0.15$~dex over 1~dex in stellar surface density. Intuitively, this comes about due to local gas being generally subject to a higher level of enrichment in regions of higher stellar density. Ref100  shows a shallower relation between gas-phase metallicity and $\Sigma_\star$ than is inferred from the data. This is likely related to the integrated MZR in EAGLE being flatter than is inferred observationally \citep{Schaye15}, explored further in \S\ref{sec:mdep}. 

By $\Sigma_\star \approx 10^{2.5} {\rm M_\odot \; pc^{-2}}$, the observed relations flatten significantly, such that the gas metallicity becomes independent of, or even anti-correlates with, $\Sigma_\star$. A turn-over is measured in the Ref100 relation, but at higher $\Sigma_\star $ than is fully captured in the data. In observational studies, the presence of a plateau in the rMZR has been attributed to a flattening or drop in the gas-phase metallicity towards the innermost parts of massive galaxies \citep[e.g.][]{RosalesOrtega11, Sanchez12, RosalesOrtega12}. Saturation of O/H in gas near the old stellar centres of galaxies, or saturation in the emission lines themselves in highly enriched gas, have been posited as alternative explanations for a flattening in the rMZR \citep[e.g.][]{RosalesOrtega12}, but do not imply a turn-over.  %

The turn-over in the rMZR may be related to that observed in the integrated MZR, both observationally \citep[e.g.][]{Yates12} and in EAGLE \citep[e.g.][]{Segers16, DeRossi17}. By exploring different feedback prescriptions, \citet{DeRossi17} show that in EAGLE the turn-over in the integrated MZR is driven by AGN feedback. Unlike stellar feedback, AGN feedback is not directly associated with enrichment of the local gas. As such, AGN may drive large scale galaxy winds that remove metal-rich gas without enriching local gas further. As the high-$\Sigma_\star$ spaxels  preferentially probe the central regions of massive galaxies, where AGN are most influential, AGN feedback seems a likely cause of the rMZR turnover in EAGLE. In addition, \citet{DeRossi17} find that AGN induce an inversion in the correlation between integrated sSFR and metallicity at high mass, which we investigate for the resolved properties in \S\ref{sec:residuals}.  We explore the evolution of the turn-over in \S\ref{sec:evo}.

We now turn to the rMZR constructed for RecHi25 galaxies, which \citet{Schaye15} showed reproduces the integrated relation better for $M_\star < 10^{10} {\rm M_\odot}$. We find that for the  $\Sigma_\star \lesssim 10^{2.5} {\rm M_\odot \; pc^{-2}}$ range, the RecHi25 rMZR exhibits a steeper slope, in better agreement with the data. However, due to the lack of high-mass galaxies captured within the RecHi25 volume, the $\Sigma_\star$ values at or above the Ref100 turn-over are not well sampled. %

We note that the application of the fiducial (MaNGA-like) and CALIFA-like weighting schemes yield only marginal differences for the EAGLE relations. For Ref100, this may also be ascribed to the flatter than observed integrated MZR, such that placing a different emphasis on galaxies at the high-mass end makes little difference to the results. For the RecHi25 relation, this may again be attributed to a limited sampling of massive galaxies.

Another interesting point of comparison is the scatter about the trend. The shaded regions around the EAGLE relations represent the 16th to 84th percentile range, which should be comparable to the $\pm1\,\sigma$ range provided for the observational data, assuming near gaussianity. Comparing to the observations, it seems that the spread about the EAGLE trend is typically larger than observed by a factor $\approx$2-4, as is the case for the integrated relation \citep{Schaye15}. A potential cause of this is the stochastic enrichment and lack of \textit{metal diffusion} in EAGLE. Gas particles are directly enriched by individual star particles in the simulation, and metals are not exchanged between them. SPH smoothing, as is employed here, goes some way towards reducing the variance in metallicity between spaxels, but has no relation to the physical process of metal diffusion. The influence of smoothing is demonstrated  by comparing the rMZR scatter of Fig.~\ref{fig:mz} to that of a point-like particle treatment in Appendix~\ref{sec:pointlike}, where the scatter is larger still.

\subsection{Dependence on galaxy stellar mass}
\label{sec:mdep}

\begin{figure*}
	\includegraphics[width=\textwidth]{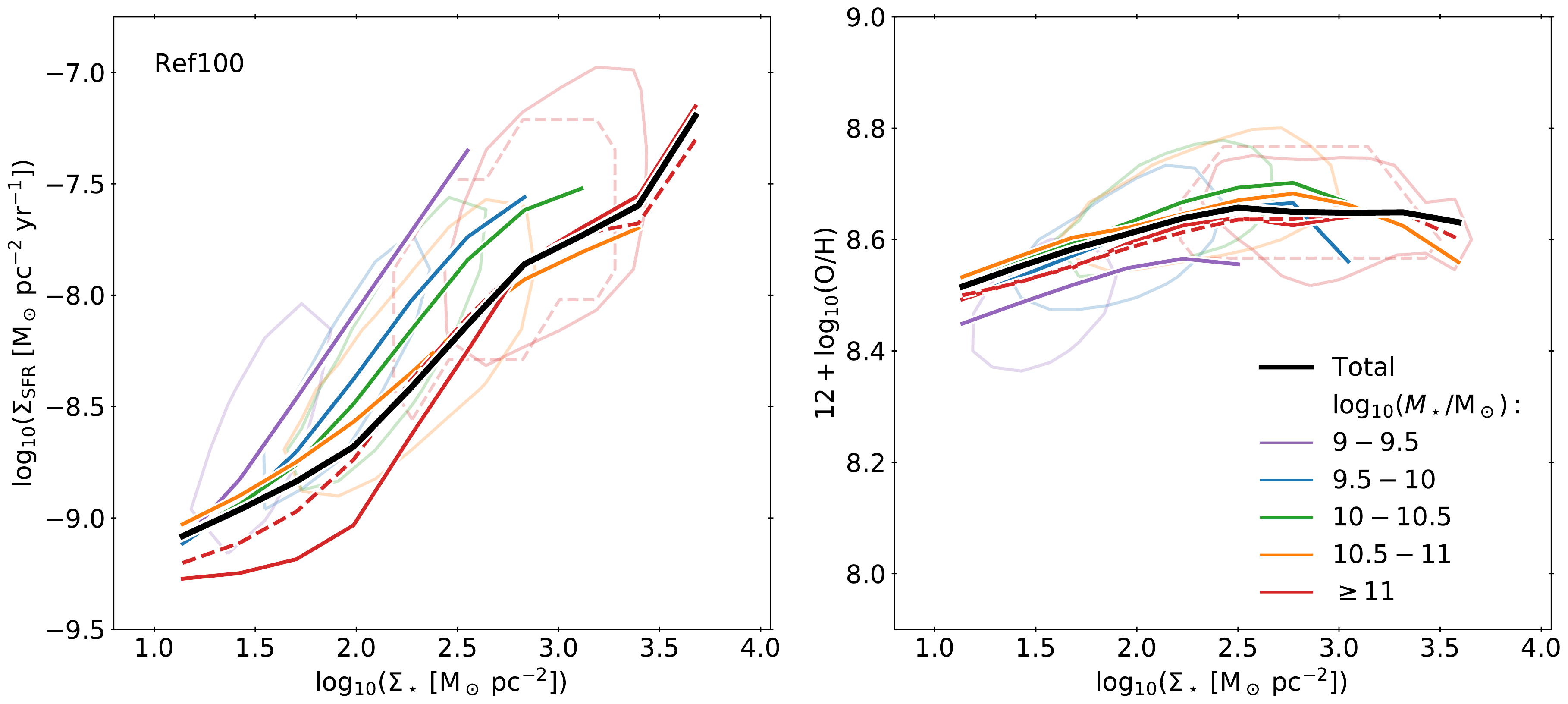}
    \caption{The spatially resolved star-forming main sequence (left, as Fig.~\ref{fig:sfms}) and stellar mass gas-phase metallicity relation (right, as Fig.~\ref{fig:mz}) for the Ref100 simulation at $z=0.1$, split into contiguous bins of integrated stellar mass. Coloured contours enclose 80\% of the weighted spaxel mass in each $M_\star$ bin, with the median relation indicated by the coloured lines. Solid lines and contours indicate the fiducial MaNGA-like spaxel weighting scheme, with dashes indicating the CALIFA weighting for the highest $M_\star$ bin. Unsurprisingly, we see that the higher mass galaxies account for the higher stellar surface density spaxels. We see that rSFMS relation in particular varies significantly with the integrated stellar mass of the contributing galaxies.}
    \label{fig:splitmass}
\end{figure*}

A clear indication that a resolved relation is  more fundamental than its integrated counterpart would be if the resolved relation does not change with the integrated properties of galaxies. Fig.~\ref{fig:splitmass} shows how the Ref-100 relations of Figs.~\ref{fig:sfms} and \ref{fig:mz} break down by the stellar mass of the contributing galaxies. Splitting Ref-100 galaxies into contiguous bins of $\log_{10}(M_\star / {\rm M_\odot})$, we plot the weighted median trends (thick coloured lines) for each mass bin, as well as the contour enclosing 80\%{} of the weighted total stellar mass (thin coloured lines). The overall weighted median trends of  Figs.~\ref{fig:sfms} and \ref{fig:mz}  are plotted for comparison (black lines).%

For the rSFMS (left panel) we first consider the weighted median lines for each mass range. We see that rather than sampling different $\Sigma_\star$ regimes of a common trend, distinct trends emerge for differing $M_\star$ bins. For $\log_{10}(M_\star/{\rm M_\odot}) < 11$, we find the trend becomes shallower with increasing  $M_\star$, with the logarithmic slope varying from $n \approx 1$ at $9 < \log_{10}(M_\star/{\rm M_\odot}) < 10$ to $n\approx 0.6$ at $10 \leq \log_{10}(M_\star/{\rm M_\odot}) < 11$. At fixed $\Sigma_\star$, $\Sigma_{\rm SFR}$ decreases with $M_\star$.

The highest $M_\star$ bin deviates somewhat from the trends observed at lower $M_\star$. At $\Sigma_\star < 10^{2}  {\rm M_\odot pc^{-2}}$, the relation is significantly flatter, and normalised 0.2-0.3~dex lower than for other mass bins. At $\Sigma_\star > 10^{2}  {\rm M_\odot pc^{-2}}$, $\Sigma_{\rm SFR}$ rises steeply, with a slope similar to that of $9 < \log_{10}(M_\star/{\rm M_\odot}) < 10$ galaxies over this $\Sigma_\star$ range.

The significant $M_\star$ dependence of the rSFMS will naturally lead to differing median trends for different mass selections. However, in Fig.~\ref{fig:sfms} it was demonstrated that the two weighting schemes yield only marginal differences. Applying a CALIFA-like weighting scheme to galaxies within each $M_\star$ bin below $\log_{10}(M_\star/{\rm M_\odot})$ yields small differences in the rSFMS with respect to our fiducial (MaNGA-like) weighting scheme. For the highest mass bin, the difference is more significant, and we show the CALIFA-like weighting using a dashed line and contour. We see that this scheme yields a trend that deviates less from the relations at lower $M_\star$, with a  steeper slope at $\Sigma_\star < 10^2\ {\rm M_\odot\ pc^{-2}}$, and a shallower slope at higher $\Sigma_\star$. These differences can be attributed to the higher weighting of the most massive galaxies in the MaNGA-like scheme relative to the CALIFA-like scheme.

It is interesting to consider why the most massive galaxies might diverge from the median relation. The star forming gas morphologies of the most massive galaxies are distinctly clumpy, shown in appendix Fig.~\ref{fig:sfmorph}, which may be indicative of a different mode of star formation in these systems. Differing star formation efficiencies may be influenced by the reservoir of gas having to build up between powerful AGN events. The steep increase in galaxy sizes for the most massive galaxies, for both the observed and EAGLE populations \citep{Furlong17}, could also play a role, with more high $\Sigma_\star$ spaxels located in the outer parts of galaxies.  %

The finding of some $M_\star$ dependence in the EAGLE rSFMS is notable. \citet{CanoDiaz16}  find no clear $M_\star$ dependence in the $\Sigma_\star-\Sigma_{\rm SFR}$ relation for CALIFA galaxies. \citet{Pan18} find some trends between $M_\star$ and the $\Sigma_\star-\Sigma_{\rm H\alpha}$ relation, exhibiting a similar flattening in the relation at higher $M_\star$, but attribute this to the influence of ionisation mechanisms other than star formation. However, other studies \citep[e.g.][]{GonzalezDelgado16, Medling18} identify strong trends between $\Sigma_\star-\Sigma_{\rm SFR}$ and galaxy morphology, a property which itself exhibits a strong trend with $M_\star$ \citep[e.g.][]{Kelvin14}.  However, it is 
unclear whether the $M_\star$ trend found in EAGLE would be recoverable in the data, given the uncertainties and contamination by ionisation not associated with star formation. We see from the 80\% contours that these trends may also be difficult to detect considering that different $M_\star$ bins contribute in different ranges of $\Sigma_\star$.

Turning now to the $\Sigma_\star$-${\rm O/H}$ relation (right panel), we again see systematic differences between $M_\star$ bins, but these are much more subtle. In the $1 \lesssim \log_{10}\Sigma_\star/ ({\rm M_\odot \; pc^{-2}}) \lesssim 2$ range, star-forming gas in spaxels from galaxies of $9 < \log_{10}(M_\star/{\rm M_\odot}) < 10$  are slightly ($\lessapprox 0.1$~dex) less enriched than those of $10 < \log_{10}(M_\star/{\rm M_\odot}) < 11$ galaxies. This indicates a certain level of \textit{non-local} chemical evolution, i.e. the metals produced by stellar populations may have enhanced gas outside of the spaxels in which they reside. As spaxels in more massive galaxies are biased to higher $\Sigma_\star$, this can lead to higher-metallicity gas at fixed $\Sigma_\star$, with additional metals contributed by regions of greater stellar density. Again, the most massive galaxies ($\log_{10}(M_\star/{\rm M_\odot}) \geq 11$) show a distinct behaviour, with systematically lower O/H values similar to that of the $9 < \log_{10}(M_\star/{\rm M_\odot}) < 9.5$ bin at low $\Sigma_\star$. However, we see a much less dramatic divergence for the most massive galaxies in the rMZR compared to the rSFMS, reflected in the similarity between trends where the MaNGA-like and CALIFA-like weighting schemes are applied.

The contours clearly display how different $M_\star$ bins sample different $\Sigma_\star$ regimes of a consistent (within 0.1~dex) global relation in EAGLE. This suggests that the local relation is more fundamental, where the integrated metallicities largely emerge from trends in enrichment on local scales. \citet{BarreraBallesteros16} find a similar result for the CALIFA sample, though they find a significantly steeper rMZR overall, as was seen in Fig.~\ref{fig:mz}. The sampling of a common rMZR also suggests why different weighting schemes make little difference to the global rMZR trend of Fig.~\ref{fig:mz}. %

\subsection{Comparing the rSFMS and rMZR residuals}
\label{sec:residuals}

To further investigate the connection between the spatially resolved $\Sigma_\star$, $\Sigma_{\rm SFR}$ and ${\rm O/H}$ values, we now investigate how the residuals of the rSFMS relate to the residuals of the rMZR.

\begin{figure*}
	\includegraphics[width=0.48\textwidth]{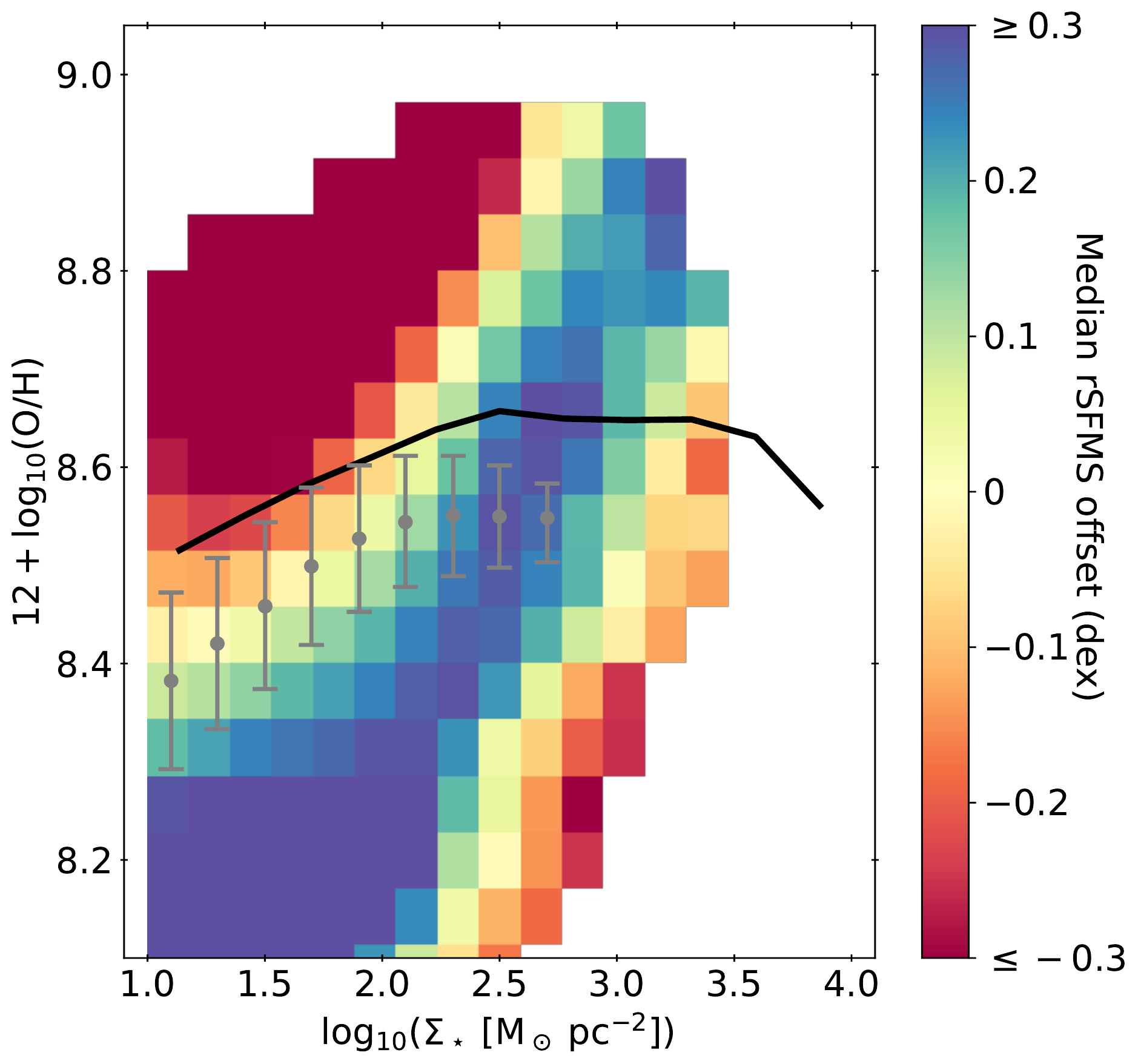}
\includegraphics[width=0.48\textwidth]{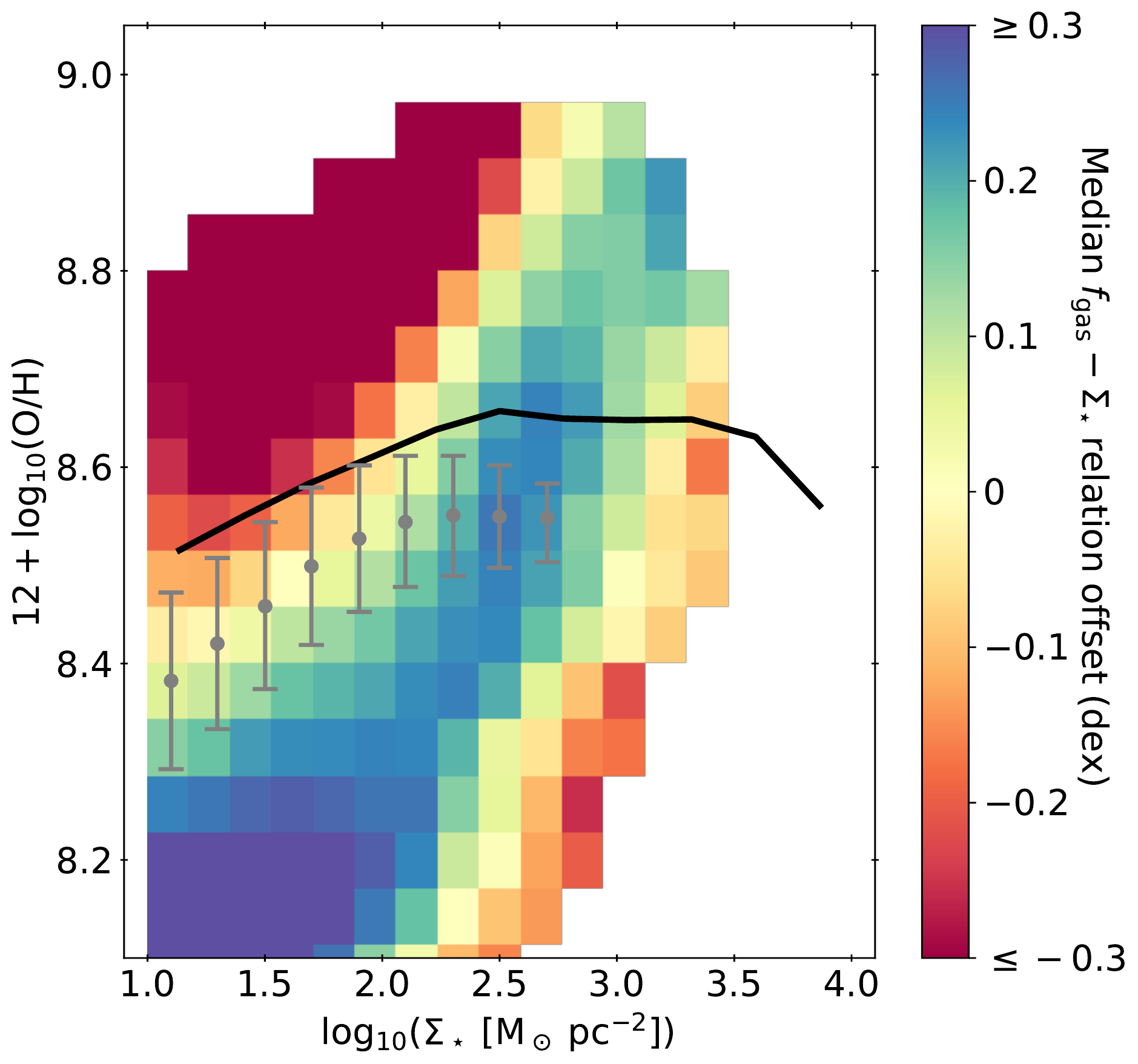}
    \caption{Relation between the residuals of the Ref100 resolved MZR relation of Fig.~\ref{fig:mz} and the median residual of the rSFMS of Fig.~\ref{fig:sfms} (colour scale in left panel) and the median resolved $f_{\rm gas}$-$\Sigma_\star$ relation (colour scale in right panel). Coloured bins are shown where more than 100 distinct galaxies contribute spaxels. The Ref100 fiducially weighted median relation (black lines) and MaNGA data of \citet{BarreraBallesteros16} (grey points) are plotted as in Fig.~\ref{fig:mz}. The residuals anti-correlate below the turnover in the rMZR, but this relationship inverts above the turnover, and displays a non-monotonic relationship around the turnover at $\Sigma_\star \sim 10^{2.5} \; {\rm M_\odot} \; {\rm pc}^{-2}$.}
    \label{fig:residuals}
\end{figure*}

In Fig.~\ref{fig:residuals}, we show the rMZR and as a thick black line, as in Fig.~\ref{fig:mz}. In the left panel, the colour scale indicates the median offset of spaxels from the rSFMS relation (see  Fig.~\ref{fig:sfms}). We compute the residuals from the interpolated  median relation shown in Fig.~\ref{fig:sfms}. The clear vertical colour gradients in Fig.~\ref{fig:residuals} imply that the residuals in the rSFMS and rMZR are strongly related. For spaxels with $\Sigma_\star \lesssim 10^{2} \; {\rm M_\odot} \; {\rm pc}^{-2}$ the residuals of the rSFMS and rMZR \textit{anti-correlate}, i.e. spaxels with higher $\Sigma_{\rm SFR}$ for their $\Sigma_\star$ tend to exhibit lower ${\rm O/H}$ and vice-versa. This is also seen in the integrated relations observationally \citep[e.g.][]{Yates12}. 

At $\Sigma_\star \gtrsim 10^{3} \; {\rm M_\odot} \; {\rm pc}^{-2}$ (i.e. above the turn-over in the rMZR) the relationship inverts, such that spaxels with lower than average star formation rates typically also have lower than average metallicities for a given $\Sigma_\star$. Interestingly, in the  $10^{2} < \Sigma_\star/({\rm M_\odot} \; {\rm pc}^{-2}) < 10^{3}$ range the relationship remains strong, but is non-monotonic, such that the most and least enriched spaxels both have relatively low $\Sigma_{\rm SFR}$. 

An analogous inversion of the relationship between \textit{integrated} metallicities and SFRs of massive galaxies has been inferred observationally \citep[e.g.][]{Yates12}. Using feedback variations in EAGLE, \citet{DeRossi17} find that this effect is driven by AGN feedback. As described in \S\ref{sec:mzr}, stellar feedback processes disrupting star formation are coincident with the enrichment of local gas by stellar winds and SNa ejecta, whereas AGN feedback events do not enhance gas metallicity. Thus, the sub-rSFMS spaxels at $\Sigma_\star \sim 10^{2.5} \; {\rm M_\odot} \; {\rm pc}^{-2}$ with high and low O/H  likely correspond to regions where star formation is suppressed by stellar and AGN feedback respectively.

The right hand panel of Fig.~\ref{fig:residuals} shows that the trends with the residuals of $\Sigma_{\rm SFR}$ are closely mirrored by the trends with the residuals of the $f_{\rm gas}(\Sigma_\star)$ relation (where $f_{\rm gas}$ is the baryon fraction in star-forming gas),  suggesting that it is the gas fraction that largely drives the scatter in the rSFMS. this mirrors the findings of \citet{Lagos16} and \citet{DeRossi17} for the integrated EAGLE relations.

\section{Evolution of scaling relations}
\label{sec:evo}

\begin{figure}
	\includegraphics[width=\columnwidth]{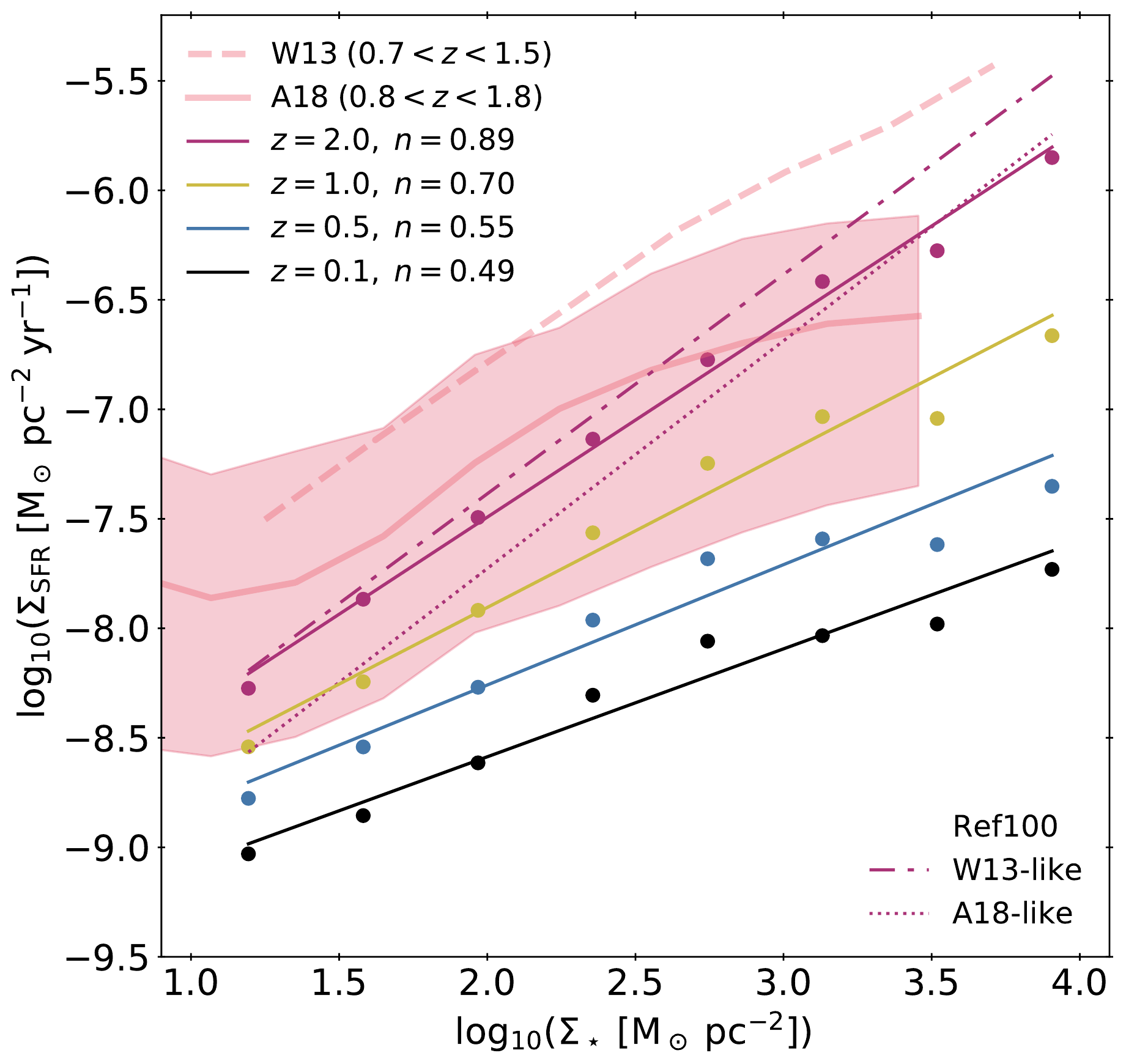}
    \caption{Evolution of the rSFMS for Ref-100. \textit{Coloured points} indicate the median values for each redshift, with \textit{thin solid lines} indicating power-law fits (the power-law index is noted in the legend). The relations are constructed for an unweighted, volume-limited sample of galaxies with $M_\star \geq 10^{10} {\rm M_\odot}$, distinct from the weighting scheme used for Fig.~\ref{fig:sfms} (see text for details). The higher-redshift data of \citetalias{Wuyts13} and \citetalias{Abdurrouf18} is included for $z \approx 1-2$, coloured to reflect the comparable redshift range of EAGLE. The galaxy selection  of \citetalias{Wuyts13} and \citetalias{Abdurrouf18} are roughly imitated to produce alternative $z=2$ EAGLE relations, and their respective power law fits are plotted using \textit{dash-dotted} and \textit{dotted} lines (see text for details). The shaded pink region indicates the 1$\sigma$ scatter on the \citetalias{Abdurrouf18} data. Both the slope and normalisation of the resolved SFMS increase with redshift.}
    \label{fig:sfmsevo}
\end{figure}

\begin{figure*}
	\includegraphics[width=0.95\textwidth]{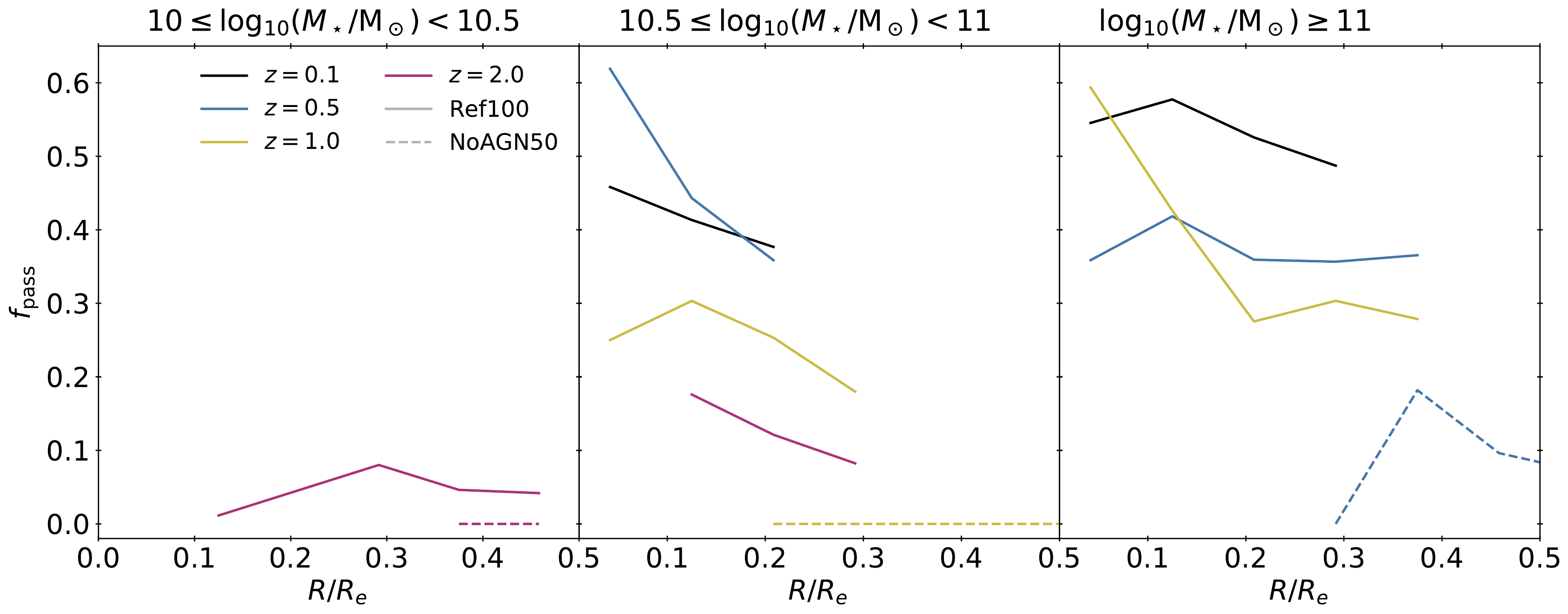}
    \caption{The evolving overall passive fraction ($f_{\rm pass}$, where $\Sigma_{\rm SFR}/\Sigma_\star > 10^{-11}$) profiles for galaxies, using the same redshift values as Fig.~\ref{fig:sfmsevo}. \textit{Solid lines} denote Ref-100, while \textit{dashed lines} represent the NoAGN50 simulation. The radii of spaxels are computed for each galaxy in terms of the projected half-mass radius, $R_e$. We only plot the profiles where more than 5 distinct galaxies contribute, and the median $\Sigma_\star$ profile exceeds $\Sigma_{\star, {\rm T}} = 10^{2.86} {\rm M_\odot \; pc^{-2} }$, where gas is deemed resolved at the $\Sigma_{\rm SFR}$ threshold for passivity (see Appendix~\ref{sec:pfrac} for details).}%
    \label{fig:ssfrprof}
\end{figure*}

The evolution of spatially resolved scaling relations can provide further clues as to how they manifest the integrated relations, and whether resolved relations are more fundamental. For example, evolution of the integrated relations could coincide with evolution of the resolved relations, or simply reflect that galaxies at different redshifts sample different parts of an unevolving resolved relation (e.g. higher $\Sigma_\star$ at higher redshift for a fixed galaxy $M_\star$). 

While observed scaling relations will probe different scales and suffer stronger selection effects at higher redshifts, we aim to isolate the physical evolution in these relations. To this end, we keep a consistent imaging methodology; we dispense with the weighting schemes used in previous sections (see section~\ref{sec:galsel}), with the same $\approx1$~kpc spaxel scale, face-on orientation and spaxel selection described in \S~\ref{sec:spaxsel} described in \S~\ref{sec:galsel}. However, we also restrict our sample to galaxies of $M_\star \geq 10^{10} {\rm M_\odot}$, in order to highlight physical evolution in an $M_\star$ regime that is captured by the high redshift data we compare to. This scheme provides a volume-limited galaxy selection down to $M_\star \geq 10^{10} {\rm M_\odot}$, which we analyse for the four simulation outputs at $z=0.1$, 0.5, 1 and 2. In addition, we make use of the standard resolution 50$^3$~Mpc$^3$ EAGLE simulation where AGN feedback is switched off (NoAGN50) to probe the evolving influence of AGN feedback in some of the following plots, employing the same galaxy and spaxel selection.

\subsection{Evolution of the rSFMS}
\label{sec:sfmsev}
We first consider the rSFMS in the Ref-100 simulation. Fig.~\ref{fig:sfmsevo} shows the $z=0.1$ median as in Fig.~\ref{fig:sfms}, but now for an unweighted and volume-limited sample of galaxies with $M_\star \geq 10^{10} {\rm M_\odot}$, and including the $z=0.5$, 1 and 2 trends. A power law is fit to the median points. We also include the high redshift data of \citet[][hereafter \citetalias{Wuyts13}]{Wuyts13} and \citet[][hereafter \citetalias{Abdurrouf18}]{Abdurrouf18}. These studies rely on high-resolution HST imaging, and broad-band SED fitting on a pixel-by-pixel basis to derive resolved properties of galaxies at $z\approx1-2$ ($0.7 < z < 1.5$ for \citetalias{Wuyts13}, $0.8 < z < 1.8$ for \citetalias{Abdurrouf18}) on kpc scales, assuming a \citet{Chabrier03} IMF. There are  differences in the target selection of these studies; \citetalias{Wuyts13} selects galaxies of $M_\star > 10^{10} {\rm M_\odot}$ and with a ${\rm sSFR} > 1/{t_{\rm Hubble}(z)}$, while \citetalias{Abdurrouf18} selects $M_\star > 10^{10.5} {\rm M_\odot}$ face-on spiral galaxies, without an explicit sSFR cut\footnote{Note, however, that a requirement that the galaxy is detected at rest-frame NUV and FUV wavelengths is imposed in \citetalias{Abdurrouf18}.}. %

Comparing to the $z \approx 1$-2 data of \citetalias{Wuyts13}, we find significant discrepancy between the observationally inferred median trends and both those predicted by EAGLE and observed by and \citetalias{Abdurrouf18}. The \citetalias{Wuyts13} observations show $\approx 0.5\,$(0.7)~dex  higher star formation rates than the EAGLE relation at $z=2\,$(1). However the shape of the relation agrees well, exhibiting a slope intermediate between $z=1$ and 2, consistent with the redshift range spanned by the data. In addition, \citetalias{Wuyts13} note a slight break and shallower slope for high-$\Sigma_\star$, which is also seen in the EAGLE data points. Again, the strength of this break in the data is intermediate between the $z=1$ and 2 cases and occurs at a similar value of $\log_{10}\Sigma_\star /({\rm M_\star \, pc^{-2}}) \approx 2.7$. 

The relation of \citetalias{Abdurrouf18} is normalised significantly lower than that of \citetalias{Wuyts13}. This is attributed by the authors to the different galaxy selection; \citetalias{Wuyts13} select preferentially star-forming galaxies, whereas \citetalias{Abdurrouf18} select star-forming and passive galaxies. Both studies select preferentially massive galaxies, with $M_\star \geq 10^{10} {\rm M_\odot}$ and $M_\star \geq 10^{10.5} {\rm M_\odot}$ for \citetalias{Wuyts13} and \citetalias{Abdurrouf18}, respectively. To assess these selection effects in EAGLE, we plot alternative $z=2$ power law fits employing selection criteria that roughly mimic \citetalias{Wuyts13} (i.e.~$M_\star > 10^{10} { \rm M_\odot}$, ${\rm sSFR} > 2.08\times 10^{-10} {\rm yr^{-1}}$) and \citetalias{Abdurrouf18} (i.e.~$M_\star > 10^{10.5} {\rm M_\odot}$). We find that the effect of this selection is relatively small in EAGLE. The \citetalias{Abdurrouf18} relation agrees better with the EAGLE relations in terms of normalisation, but shows a stronger turnover, taking place at lower-$\Sigma_\star$ values. This leads to a significantly shallower relation for galaxies of  $\log_{10}\Sigma_\star / ({\rm M_\star \, pc^{-2}}) \gtrsim 2.5$. We also show the 1$\sigma$ scatter on the observed \citetalias{Abdurrouf18} relation  (pink shaded region), demonstrating that the variance in $\Sigma_{\rm SFR}(\Sigma_\star)$ is comparable to the level of difference between the EAGLE and \citetalias{Wuyts13} relations. 

While differences in the normalisation, slope and shape of the resolved SFMS exist between EAGLE and the data at $z \approx 1-2$, these are at a similar level to those between the two available data sets. Determining resolved properties at these redshifts is very challenging, with uncertainties in pixel-by-pixel SED fitting at these redshifts and galaxy selection effects potentially contributing significant systematic effects. It is difficult to disentangle the influence of observational systematics from true inadequacies in the EAGLE simulation. However, a well known issue with galaxy formation models, including EAGLE, is the underprediction of the observed $\dot{M}_\star(M_\star)$ relation at $z \approx 2$ \citep[e.g.][]{Weinmann12, Genel14, Henriques15, Furlong15}. A number of plausible explanations have been suggested for this discrepancy, such as shortcomings in the implementation of feedback and subsequent reincorporation times of galaxies \citep[e.g.][]{Mitchell14}, or the effect of a top-heavy IMF in highly star-forming galaxies in changing measured SFRs \citep[e.g.][]{Hayward11, Zhang18, Cowley18}. %

The EAGLE rSFMS evolves significantly with redshift, both in terms of its normalisation and slope. The typical $\Sigma_{\rm SFR}$ steepens with redshift for a fixed  $\Sigma_{\star}$ over the entire sampled range. This increase is more pronounced at high $\Sigma_{\star}$, yielding a power law slope ($n$, inset in Fig.~\ref{fig:sfmsevo}) that increases with redshift. We can compare this result with the observed evolution of the integrated main sequence using the equation of \citet{Speagle14}, derived from a compilation of observational results. Qualitatively, the observed integrated main sequence evolution is similar, with increasing normalisation and slope as a function of redshift. However, while the observed integrated main sequence slope decreases by a factor $\approx 1.5$ (0.76 to 0.52) from $z=2$ to $0.1$ \citep{Speagle14}, the slope of the EAGLE rSFMS decreases by a factor of $\approx 1.8$ (0.89 to 0.49) over the same interval. 

The more rapid evolution in the slope for the resolved main sequence could indicate a change in how star formation is distributed within galaxies as a function of redshift. In particular, this may be compatible with the concept of \textit{`inside-out'} formation: where the inner parts of galaxies evolve more rapidly than the outer parts. However, it could also be attributable to the evolving $M_\star$ distribution of galaxies, and the strong $M_\star$ dependence of the rSFMS found in EAGLE (seen in Fig.~\ref{fig:splitmass}).  

In order to test this more directly, in Fig.~\ref{fig:ssfrprof} we plot the overall passive fraction ($f_{\rm pass}$) of spaxels as a function of their radius in units of $R_e$, at each redshift and in bins of $M_\star$. These profiles are computed using both active and passive spaxels, and the projected half-mass radii, $R_e$, of \citet{Furlong17}. A spaxel is deemed passive if the specific SFR ($\Sigma_{\rm SFR}/\Sigma_\star$) is below a constant threshold of $10^{-11}$~yr$^{-1}$, equal to the integrated passivity threshold for $z=0.1$ galaxies used by \citet{Furlong15}. A subtlety of computing passive fractions is that, for a spaxel of a given $\Sigma_\star$, the passive threshold should correspond to a sufficiently resolved amount of gas to prevent $f_{\rm pass}$ from being influenced by sampling effects and shot noise. The computation of passive fractions is expounded in Appendix~\ref{sec:pfrac} where it is found that $f_{\rm pass}$ is sufficiently resolved above $\Sigma_{\star,{\rm T}} = 10^{2.86} {\rm M_\odot \; pc^{-2}}$. In Fig.~\ref{fig:ssfrprof}, we only plot profiles where the median $\Sigma_\star$ exceeds $\Sigma_{\star,{\rm T}}$, and more than 5 galaxies contribute.  

Given the stringent $\Sigma_\star$ criterion, we only show radii within $0.5 R_e$, where some $f_{\rm pass}$ values are reliable. We first focus on the $10.5 \geq \log_{10}(M_\star/{\rm M_\odot}) > 11$ range in the middle panel, where  profiles can be found for each redshift. We find that galaxies are typically more passive towards their centres. In addition, we see that for galaxies in this same $M_\star$ range, the passive fractions generally increase with redshift, with spaxels generally becoming passive earlier at shorter radii. 

While this hints towards an \textit{inside-out} picture of galaxy formation, this is not particularly compelling, due to the limited regime in which we can determine passivity on kpc scales in EAGLE. There are also exceptions, for example in the $10.5 \geq \log_{10}(M_\star/{\rm M_\odot}) > 11$ panel, the $f_{\rm pass}$ in the very centres (leftmost point) of $z=0.5$ galaxies exceeds that of $z=0.1$ galaxies (only 16 and 18 galaxies contribute to this bin, respectively). We note that in the high $\Sigma_\star$ regime probed here, the role of AGN feedback is key. This is demonstrated by the NoAGN50 profiles showing that, without AGN feedback, the spaxels in the central regions we probe are almost exclusively star-forming ($f_{\rm pass} \approx 0$). These galaxies are also more compact, such that we cannot measure the profiles down to the lowest $R/R_{e}$ values measured for at the kpc-scale resolution of our spaxels. 

It is interesting to consider the evolution of resolved star formation in light of the recent results of \citet{Starkenburg18}. They find that, contrary to what is reported observationally, $z=0$ EAGLE (and Illustris) galaxies appear to \textit{quench} in an \textit{outside-in} fashion, in the sense that the simulated sub-SFMS galaxies exhibit more centrally peaked sSFRs. This result is not necessarily in tension with an inside-out \textit{formation} of galaxies overall. Indeed, the general increase in galaxy sizes with cosmic time \citep{Furlong17} implies that galaxies must build up the majority of their stellar mass \textit{inside-out}. In fact, \citet{Clauwens17} explicitly demonstrated that EAGLE galaxies grow inside out, and may even do so somewhat more prominently than observed. Our finding that the star formation efficiency of higher $\Sigma_\star$ regions drops more rapidly with cosmic time than lower $\Sigma_\star$ regions also supports this picture. 

The $z=0.1$ $f_{\rm pass}$ profiles of Fig.~\ref{fig:ssfrprof}  provide a more direct probe of radial \textit{quenching} in EAGLE galaxies at $z=0.1$. These don't exhibit an outside-in trend, but, as noted previously, the stringent criteria to resolve the passive threshold means that this result cannot be considered representative of the overall population.

\subsection{Evolution of the rMZR}
\label{sec:mzrev}
\begin{figure}
	\includegraphics[width=\columnwidth]{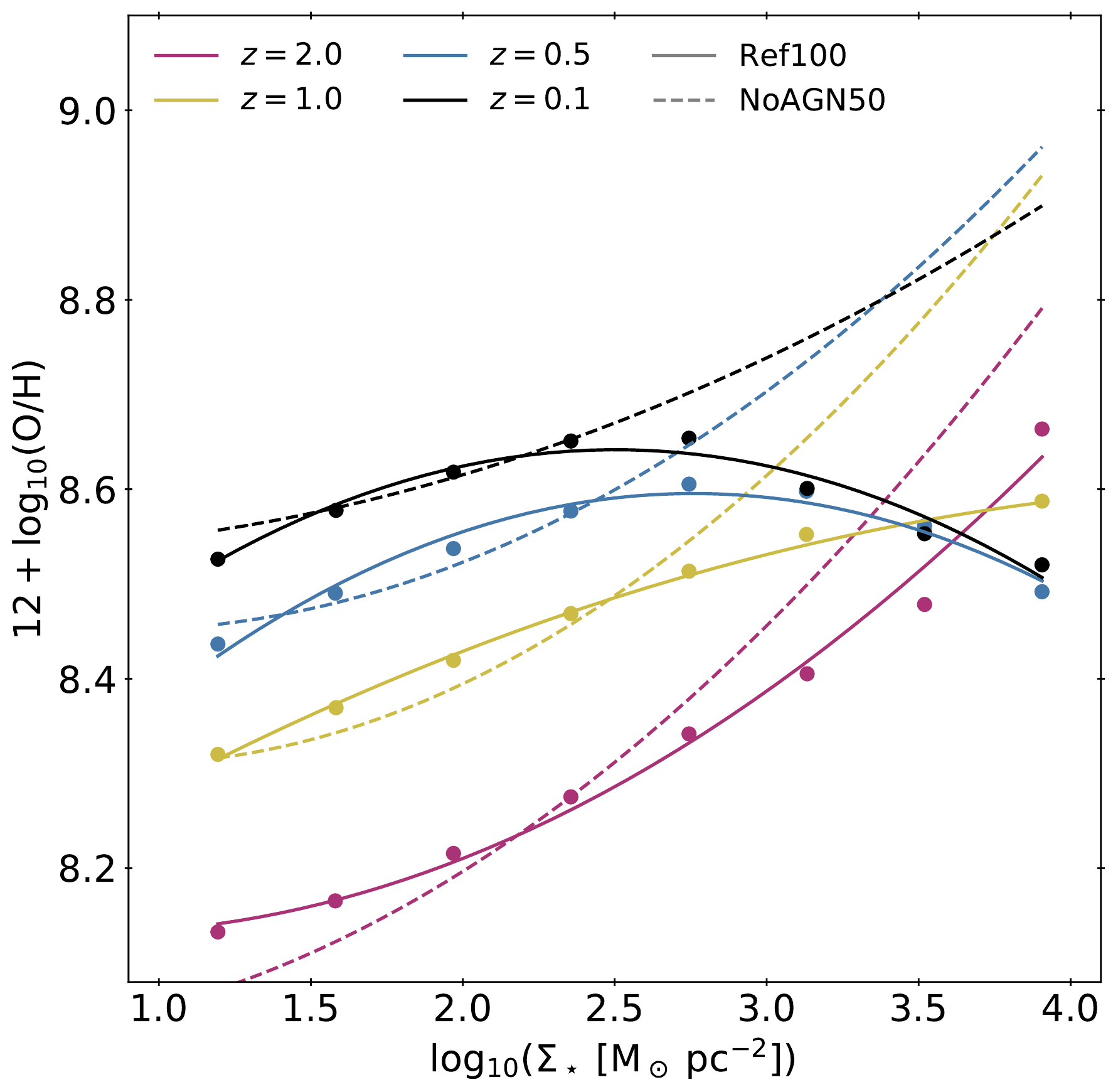}
    \caption{The median $\Sigma_\star-{\rm O/H}$ (rMZR) relation of Fig.~\ref{fig:mz}, but now plotted at multiple redshifts and constructed using an unweighted, volume-limited sample of $M_\star > 10^{10}$ galaxies used in Fig~\ref{fig:sfmsevo}. We apply a constant recalibration factor to the EAGLE data (see section~\ref{sec:zcal}). Median points are shown for the Ref100 run, while \textit{solid lines} show the best fit quadratic polynomial between $\log_{10}({\rm O/H})$ and $\log_{10}(\Sigma_\star)$ for each redshift. \textit{Dashed lines} also show quadratic fits to the NoAGN50 rMZR for comparison. There is significant evolution in the shape of the Ref100 relation as well as in the normalisation, particularly for low-$\Sigma_\star$ spaxels.}
    \label{fig:mzevol}
\end{figure}

We now explore the evolution of the $\Sigma_\star-{\rm O/H}$, or rMZR, relation. Fig.~\ref{fig:mzevol} shows the Ref100 $z=0.1$ rMZR of Fig.~\ref{fig:mz}, except now using the unweighted, volume-limited sample limited to $M_\star > 10^{10}$. The rMZR at $z=0.5$, 1 and 2 are constructed in the same way. To capture the high-$\Sigma_\star$ turnover identified at $z=0.1$, we fit a second-order polynomial in  $\log_{10}({\rm O/H})$ and $\log_{10}(\Sigma_\star)$ to the EAGLE data. Again, we concentrate on relative differences between redshifts and the shape of the relation, and not the absolute normalisation of abundance values, which have been shifted as discussed in \S~\ref{sec:zcal}.

First considering the Ref100 $\Sigma_\star-{\rm O/H}$ relation, we see remarkable evolution in the shape of the relation. At $z=2$, the relation is convex, with metallicity increasing by $0.5$~dex from $\Sigma_\star \sim 10 {\rm \; M_\odot pc^{-2}}$ to $\Sigma_\star \sim 10^4 {\rm \; M_\odot pc^{-2}}$. At $z=1$ the relation becomes shallower and close to linear with metallicity, increasing by 0.3~dex at the same stellar surface density range. For $z=0.5$ and $z=0.1$ the relation has become concave, and spans $\lesssim 0.2$~dex in abundance. While the gas-phase abundances in $\Sigma_\star \lesssim 10^{3}$ spaxels tend to decrease with redshift, at the highest $\Sigma_\star$ the metallicity remains nearly constant.%

It is notable that the gas-phase abundances evolve at fixed $\Sigma_\star$ in EAGLE. This demonstrates that the redshift dependence of the integrated MZR cannot be fully explained by an evolving galaxy population sampling an unevolving local relation. For $\Sigma_\star \lesssim 10^{3} {\rm \; M_\odot pc^{-2}}$, lower abundances at higher redshift is ascribed to increased gas fractions, higher outflow rates, and fewer prior stellar generations in galaxies. This has been demonstrated in EAGLE for integrated gas-phase abundances, which evolve roughly as observed even though a remarkably static relation between gas fraction and metallicity exists for most of cosmic time \citep{DeRossi17}.

The evolution at $\Sigma_\star > 10^{3} {\rm \; M_\odot pc^{-2}}$ is perhaps more challenging to understand. It seems likely that the inversion of the $\Sigma_\star > 10^{3} {\rm M_\odot \; pc^{-2}}$ trend at high redshift is related to AGN feedback, as such $\Sigma_\star$ values are only found in the central parts of massive galaxies. To test the influence of AGN directly, we also show the evolution of the NoAGN50 rMZR for comparison (dashed lines). We see that the NoAGN50 rMZR is similarly convex, if slightly steeper, at $z=2$. While the NoAGN50 rMZR shallows between $z=2$ and 0.1, the relation remains convex, with no evidence of a flattening or turnover. This shows explicitly that the Ref100 rMZR inversion can be attributed to AGN. %

In lieu of measurements of the rMZR in high-redshift galaxies, the evolutionary picture provided by EAGLE remains to be tested observationally.

\subsection{The resolved gas fraction-metallicity relation}
\label{sec:gfrac}

Relations that are independent of redshift may point towards fundamental aspects of galaxy evolution. A fundamental three-dimensional relation between the integrated properties of $M_\star$, $Z_{\rm gas}$ and  SFR has been identified observationally \citep[e.g.][]{LaraLopez10}, though this appears to break down at high redshift
\citep[$z \gtrsim 2$, e.g.][]{Mannucci10, Salim15}. In EAGLE, \citet{Lagos16} identify a more persistent plane relation by replacing $Z_{\rm gas}$ with the neutral gas fraction, motivated by observational trends. \citet{Matthee18} find that EAGLE predicts smaller scatter when ${\rm \upalpha}$-enhancement is used instead of metallicity. Furthermore, \citet{DeRossi17} show that in EAGLE $Z_{\rm gas}$ and gas-fraction exhibit a strong redshift-independent anti-correlation. Testing the redshift independence of the resolved $Z_{\rm gas}$-$f_{\rm gas}$ relation provides insight into whether or not the integrated relation is borne of more fundamental local relations.

\begin{figure}
	\includegraphics[width=\columnwidth]{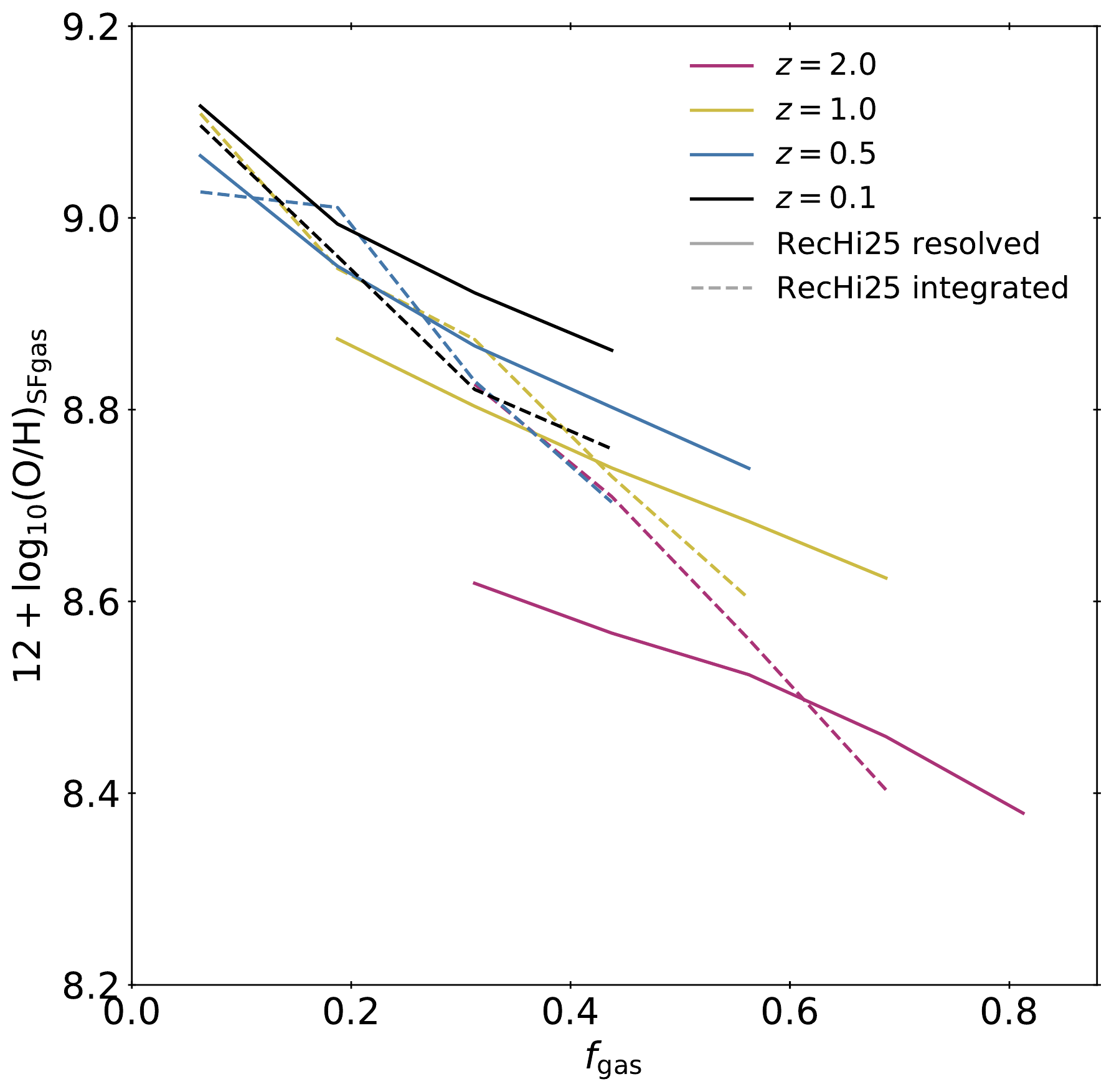}
    \caption{The spatially resolved  median gas-phase ${\rm O/H}$ as a function of $f_{\rm gas}$ relation for individual spaxels from Recal-25 galaxies, plotted for $z=0.1$, 0.5, 1 and 2 (different colours). \textit{Solid} and \textit{dashed lines} denote the resolved and integrated relations respectively. The resolved relation does not show the redshift independence of its integrated counterpart. }
    \label{fig:zfgas}
\end{figure}

We plot $Z_{\rm gas}$ as a function of the star-forming gas fraction (the ratio of star-forming gas mass to total baryonic mass), $f_{\rm gas}$, at different redshifts in Fig.~\ref{fig:zfgas} for both resolved and integrated values, such that individual kpc scale spaxels and individual galaxies contribute respectively. We construct these relations using the Recal-25 simulation,  which better reproduces observations of the integrated MZR and for which the redshift independent integrated trend is recovered \citep{DeRossi17}. 

Interestingly, the near redshift-independence exhibited by the integrated relation does not extend to the resolved relation. For spaxels with $f_{\rm gas} \approx 0.4$, for example, the median metallicity increases by $\approx0.3$~dex between $z=2$ and 0.1 for this $f_{\rm gas}$, while the integrated metallicities differ by less than 0.1~dex . The resolved relations  show a negative trend, but one that is shallower than for their integrated counterparts at each redshift. 

The anti-correlation between $Z_{\rm gas}$ and $f_{\rm gas}$ is intuitive: as the gas mass increases relative to the stellar mass, the stellar ejecta that enrich the ISM become more dilute. This scenario is true of a simple \textit{`closed-box'} \citep[e.g.][]{Schmidt63, Tinsley80} model for metal enrichment, but this is clearly not a representative model of a galaxy experiencing continuous inflows and outflows. Indeed, \citet{DeRossi17} show that the overall effective yields measured in the ISM of EAGLE galaxies are below the intrinsic stellar yields, particularly in the AGN dominated high-$M_\star$ regime, indicative of the role of gas flows. Instead, a \textit{dynamic equilibrium} model of metal enrichment \citep[e.g.][]{Erb06, Dave11}, where inflows and outflows balance due to self-regulating feedback processes, likely provides a more suitable description for EAGLE galaxies. Such a model reproduces the tight, unevolving trend between the global gas fractions and gas-phase O/H in EAGLE galaxies remarkably well \citep{DeRossi17}.  %

Neither model provides a fitting description of the evolution of the rMZR. This points to the importance of \textit{non-local} aspects of feedback and enrichment processes within galaxies. With no physical barriers separating projected 1~kpc$^2$ regions of a galaxy, stellar feedback causes enriched gas to be redistributed within a galaxy. While the galaxy as a whole may be in a state of near dynamic equilibrium, the non-local nature of enrichment and feedback tends to weaken the anti-correlation between $f_{\rm gas}$ and O/H on these scales. %

\subsection{Interpreting the evolution of resolved relations}
\label{sec:interp}

An interesting question from the analysis of Figs.~\ref{fig:sfmsevo}-\ref{fig:zfgas} is whether the mode of star formation evolves through cosmic time. The relative importance of star formation in the inner and outer parts of galaxies is one aspect of this. The shallowing of the rSFMS with decreasing redshift (Fig.~\ref{fig:sfmsevo}) indicates that galaxy centres were more star forming in the past relative to outer regions, building up their bulges. Tentative evidence of this is seen in the Fig.~\ref{fig:ssfrprof}, where the central regions of galaxies are found to become passive earlier on in cosmic history. We note that this is not necessarily in contradiction with the recent findings of outside-in \textit{quenching} in $z=0$ EAGLE galaxies by \citet{Starkenburg18}. It may be enlightening in future work to test if imitating optical observations \citep[as in][]{Trayford17} and their selection effects, including pollution by ionisation sources not associated with star formation, allows the same qualitative results to be recovered.  

The evolution of the rMZR (Fig.~\ref{fig:mzevol}) provides further insight, where the steep slope in the highest density regions ($\Sigma_\star >  10^{3} {\rm M_\odot \; pc^{-2}}$) at $z=2$ inverts to become a distinct downturn by $z=0.1$. These high-density spaxels preferentially probe the inner regions of massive galaxies, a regime where the behaviour of AGN feedback is particularly important. While the turnover in the $z=0.1$ trend is indicative of AGN efficiently removing gas from galaxy centres (see section~\ref{sec:mzr}), the convex trend at $z=2$ suggests ongoing star formation in the centres of massive galaxies, despite the presence of AGN. This can be ascribed to a smaller fraction of super-massive black holes having grown to the \textit{AGN regulated} phase by this time, where AGN feedback can efficiently disrupt star formation \citep{McAlpine18}.

\section{Summary \& Conclusions}
\label{sec:summary}

We have produced maps of spatially resolved physical properties for the virtual galaxies formed in the EAGLE simulation, and presented resolved scaling relations between various properties. In particular, we focused on the resolved `\textit{star-forming main sequence}' (rSFMS) and resolved gas-phase mass-metallicity relations (rMZR) on 1~kpc scales. We assessed the simulated scaling relations by comparing to data inferred from local IFU surveys \citep{Sanchez12,Bundy15}, and high redshift imaging \citep{Wuyts13,Abdurrouf18}. For the SFMS we considered both the shape and normalisation of the relation, but for the resolved MZR we ignored the normalisation, due to the large systematic uncertainty in the absolute normalisation of metallicity values. \\

We first concentrated on the local ($z=0.1$) relations, which were constructed from an appropriately weighted galaxy sample for comparison to the local MaNGA and CALIFA surveys. Our general conclusions are as follows:
\begin{itemize}
\item The rSFMS slope agrees remarkably well with observations for both the standard resolution fiducial 100$^3$~cMpc$^3$ volume (Ref100) and the high resolution 25$^3$~cMpc$^3$ volume with recalibrated subgrid parameters (RecHi25).%
\item The normalisation of the EAGLE rSFMS relations show a $\approx-0.15$~dex offset from the observed relation for the fiducial MaNGA weighting scheme. This is consistent with the $\approx-0.2$~dex offset from observations found for the integrated SFMS \citep{Furlong15}, given the observational uncertainties. This suggests that the $z=0.1$ integrated and resolved main sequence relations can be brought into simultaneous agreement with the observations by changing the absolute normalisation of SFRs alone. We note that there is uncertainty in the absolute normalisation of SFRs observationally, with some studies finding close agreement with EAGLE \citep[e.g.][]{Chang15}.
\item The shape of the rMZR relation also generally follows that of the data (see Fig.~\ref{fig:mz}), with the Ref100 simulation in particular exhibiting a peak at a similar characteristic stellar surface density as seen in the data, $\Sigma_\star \sim 10^{2.5} {\rm M_\odot \; pc^{-2}}$. Again, imitating the selection function of CALIFA instead of MaNGA induces only marginal differences in the rMZR. %
\item The rSFMS and rMZR exhibit qualitatively different dependences on the $M_\star$ of their host galaxies. In the mass range $9 \leq \log_{10}(M_\star/{\rm M_\odot}) < 11$, the rSFMS slope depends on the $M_\star$ of galaxies from which it is constructed, with a best-fit power law index decreasing from $n\approx1$ to $n\approx0.7$ as $M_\star$ increases by 2~dex. Higher-mass galaxies deviate from this trend. The rMZR shows a much weaker $M_\star$ dependence, varying by only $0.1$~dex between $M_\star$ bins (Fig.~\ref{fig:splitmass}). 
\item The residuals of the rSFMS and rMZR are strongly related (Fig.~\ref{fig:residuals}); the correlation is negative at low $\Sigma_\star$ ($\Sigma_\star < 10^2 {\rm M_\odot \; pc^{-2}}$) and positive at high $\Sigma_\star$ ($\Sigma_\star > 10^3 {\rm M_\odot \; pc^{-2}}$). In the surface density range, the relation is strong but \textit{non-monotonic} with the spaxels lying above and below the rMZR both having preferentially low $\Sigma_{\rm SFR}$. The inversion of this relation is ascribed to a transition between the regimes where feedback is predominately driven by SF and AGN. 
\end{itemize}

We then considered how EAGLE predicts these relations to evolve, using an unweighted, volume-limited sample of galaxies with $M_\star > 10^{10} {\rm M_\odot}$. We find that: 
\begin{itemize}
\item The rSFMS evolves, shifting to higher $\Sigma_{\rm SFR}$ values and becoming steeper with redshift. The observed rSFMS is normalised significantly higher at $z\approx2$, as was also found for the integrated SFMS \citep{Furlong15}, but appears consistent within the large error range (Fig.~\ref{fig:sfmsevo}). 
\item The rMZR also shows strong evolution, both in terms of its shape and normalisation. The resolved MZR evolves from a steep, convex relation at $z=2$, to the shallower concave relation found at $z=0.1$.  While the rMZR exhibits a positive trend at all redshifts for stellar surface densities of $\Sigma_\star < 10^{2.5}$~M$_\odot$~pc$^{-2}$, at higher $\Sigma_\star$, the rMZR evolves from a positive to negative trend between $z=2$ and $z=0.1$  (Fig.~\ref{fig:mzevol}). We demonstrated that the turn-over in the low-redshift rMZR is induced by AGN feedback, showing that the relation found for the NoAGN50 simulation volume, where AGN feedback is not included, remains convex throughout cosmic time.  %
\item The redshift independence of the EAGLE integrated $\log_{10}({\rm O/H})$-$f_{\rm gas}$ relation found by \citet{DeRossi17} is not exhibited by its resolved counterpart (Fig.~\ref{fig:zfgas}). This is attributed to the \textit{non-local} processes that emerge within EAGLE galaxies. The enrichment of the gas probed by a spaxel cannot be attributed solely to the stars within that spaxel, because gas and metals move between individual kpc-scale regions.%
\end{itemize}

Altogether, we find that while the existence of spatially resolved scaling relations indicates that physical processes such as star formation, chemical enrichment and feedback also depend and operate on small scales, these resolved relations are not more `\textit{fundamental}'. The fact that the resolved scaling relations evolve, implies that the evolution of their integrated counterparts is not merely the result of higher-redshift galaxies having different surface density profiles. Indeed, the evolution of galaxies depends on both local and galaxy-wide processes.

While mapping physical properties of EAGLE galaxies appears to show resolved scaling relations in reasonable agreement with observations, we emphasise that observationally these scaling relations must be derived from observable properties. A more direct comparison would be to use virtual observations of EAGLE galaxies \citep{Camps16,Trayford17} to derive resolved properties using the same approach, potentially building in the same assumptions and systematic effects present in the data. %

\section*{Acknowledgements}

We thank Sebastian S\'{a}nchez for useful insight into the calibration of metallicity measurements from observations. This study made use of the publicly available {\tt py-sphviewer} code \citep{BenitezLambay15}. Our analysis was carried out using the DiRAC Data Centric system at Durham University, operated by the Institute for Computational Cosmology on behalf of the STFC DiRAC HPC Facility ({\tt www.dirac.ac.uk}). This equipment was funded by BIS National E-infrastructure capital grant ST/K00042X/1, STFC  capital  grant  ST/H008519/1,  and  STFC  DiRAC Operations  grant  ST/K003267/1  and  Durham  University. DiRAC is part of the National E-Infrastructure.

\bibliographystyle{mnras}
\bibliography{refs}

\appendix
\section{Convergence Testing}
\label{sec:convergence}

\begin{figure*}
	\includegraphics[width=0.48\textwidth]{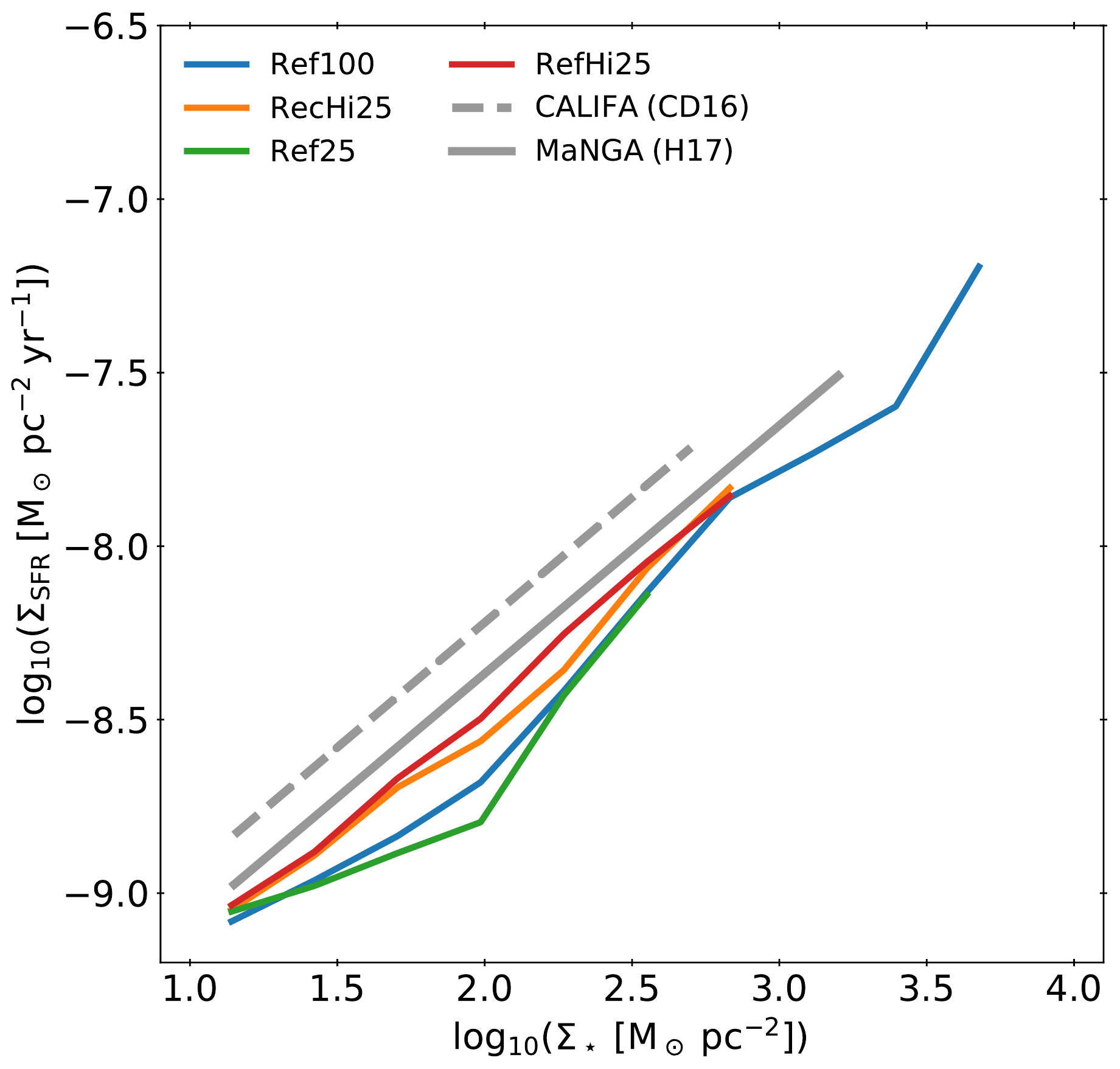}
\includegraphics[width=0.48\textwidth]{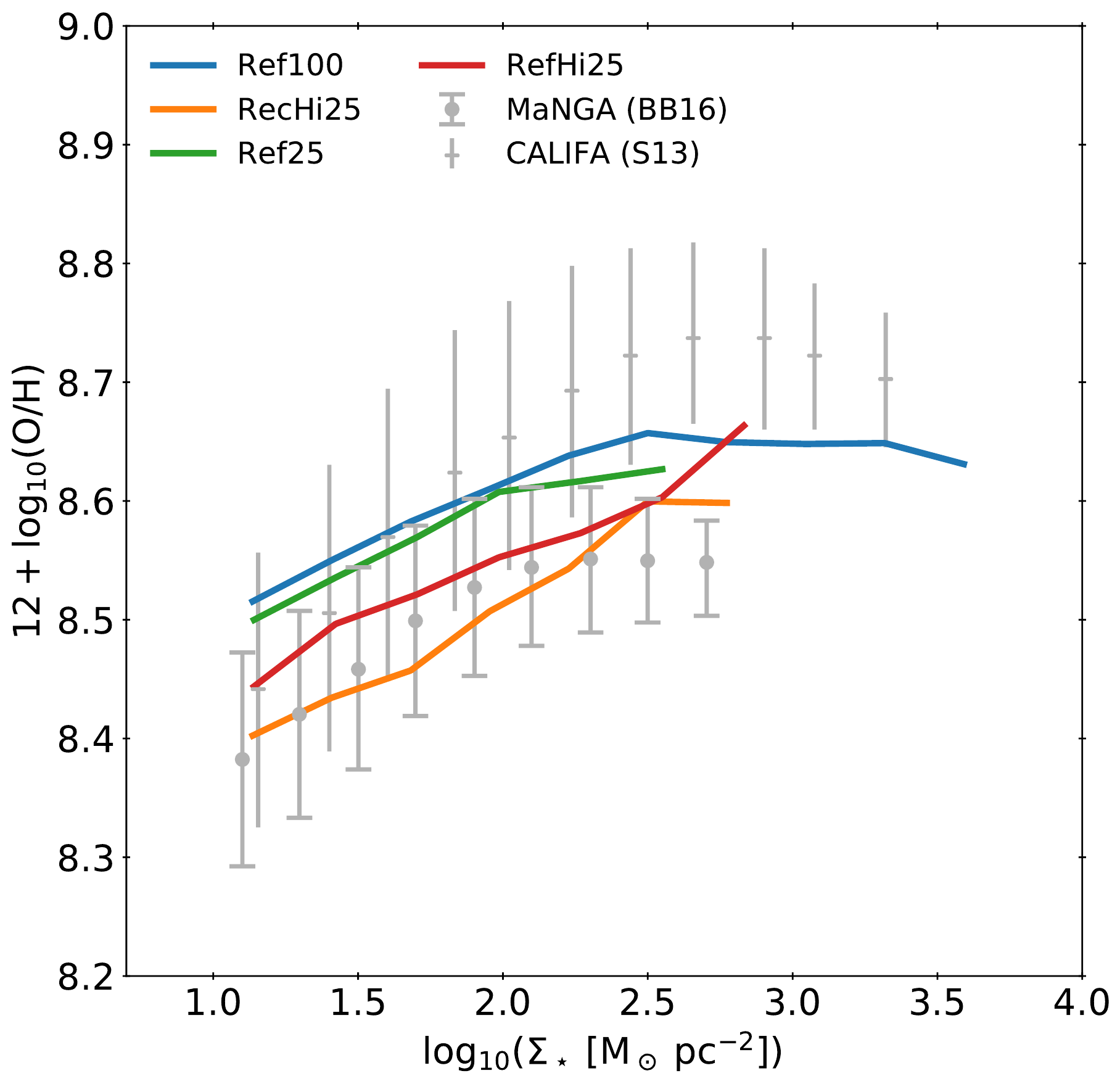}
    \caption{The rSFMS (left panel) and rMZR  (right panel) for a variety of different simulations, in order to test the influence of numerical convergence and volume selection effects. The Ref100-Ref25 comparison shows the influence of volume effects, whereas the Ref25-RefHi25 and Ref25-RecHi25 show the \textit{`strong'} and \textit{`weak'} convergence properties, respectively \citep[see][]{Schaye15}. We find that the rSFMS converges reasonably well, but the rMZR is slightly steeper at higher resolution, in better agreement with the data.}
    \label{fig:convergence}
\end{figure*}

We test the convergence of the two scaling relations plotted in Figs.~\ref{fig:sfms} and \ref{fig:mz} in Fig.~\ref{fig:convergence}. We show the weighed median relations for the simulation volumes used in this study (Ref100 and RecHi25), as well as the RefL025N0376 (Ref25) and RefL025N0752 (RefHi25) volumes for additional convergence testing. The Ref25 simulation uses the reference model and standard resolution, but differs from Ref100 by sampling a smaller volume. The RefHi25 volume is the same volume and resolution as RecHi25, but using the reference subgrid physics parameter values.    

We see that for the rSFMS (left panel) the effects of volume sampling and resolution have little influence on the relation over the $\Sigma_\star$ range covered by all the simulations ($\Sigma_\star \lesssim 10^{2.5} \; {\rm M_\odot} \; {\rm pc}^{-2}$), suggesting that this relation converges well.

Comparing the rMZR of Ref100 and Ref25 (right panel), we see that the shape of the relation is preserved reasonably well for the same model and resolution sampling a smaller volume. The turnover is exhibited  at the same value of $\Sigma_\star \sim 10^{2.5} \; {\rm M_\odot} \; {\rm pc}^{-2}$, but plateaus at lower abundances in the smaller volume. The close rMZR agreement between Ref100 and Ref25 over the sampled $\Sigma_\star$ range suggests that the influence of volume effects are relatively small. The higher resolution runs display systematically lower abundances (by $\lesssim 0.1$~dex), pointing to a level of numerical influence on the resolved ${\rm O/H}$ values, though we do not consider normalisation of abundances in this work (see \S\ref{sec:zcal} for details). Instead, we consider only the relative abundances, expressed through the slope of the rMZR. We find that at $\Sigma_\star <  10^{2} \; {\rm M_\odot} \; {\rm pc}^{-2}$ the rMZR slopes do not show a systematic difference with resolution, with RefHi25 and RecHi25 respectively exhibiting subtly shallower and steeper slopes than that of Ref100.  At $\Sigma_\star >  10^{2} \; {\rm M_\odot} \; {\rm pc}^{-2}$, the comparison becomes less appropriate, as $\Sigma_\star$ values are increasingly biased towards high $M_\star$ galaxies missing from the 25$^3$~cMpc$^3$ volumes. As a result, we can infer little about the convergence properties of the rMZR peak.

\section{The influence of intrinsic smoothing}
\label{sec:pointlike}

Material in SPH simulations, such as EAGLE, is discretised by mass into \textit{particles}. The spatial extent of a particle is not rigidly defined. For gas particles, an intuitive size to use is that of the SPH kernel used by the simulation to compute hydrodynamical forces, and computed using the distance to the nearest neighbours. Star particles do not have a similar associated size. As a result, stellar smoothing is chosen in a somewhat \textit{ad-hoc} way, as described in \citet{Trayford17}.

In order to test the influence of this smoothing on our kpc-scale resolved scaling relations, we compare to the extreme case of \textit{no smoothing} (i.e. treating particles as point-like). In this case, particles contribute only to the spaxels they are projected onto. By also producing sets of property maps for galaxies where particles are treated as point-like, we can get an idea of how intrinsic smoothing influences our results.

\begin{figure*}
	\includegraphics[width=0.48\textwidth]{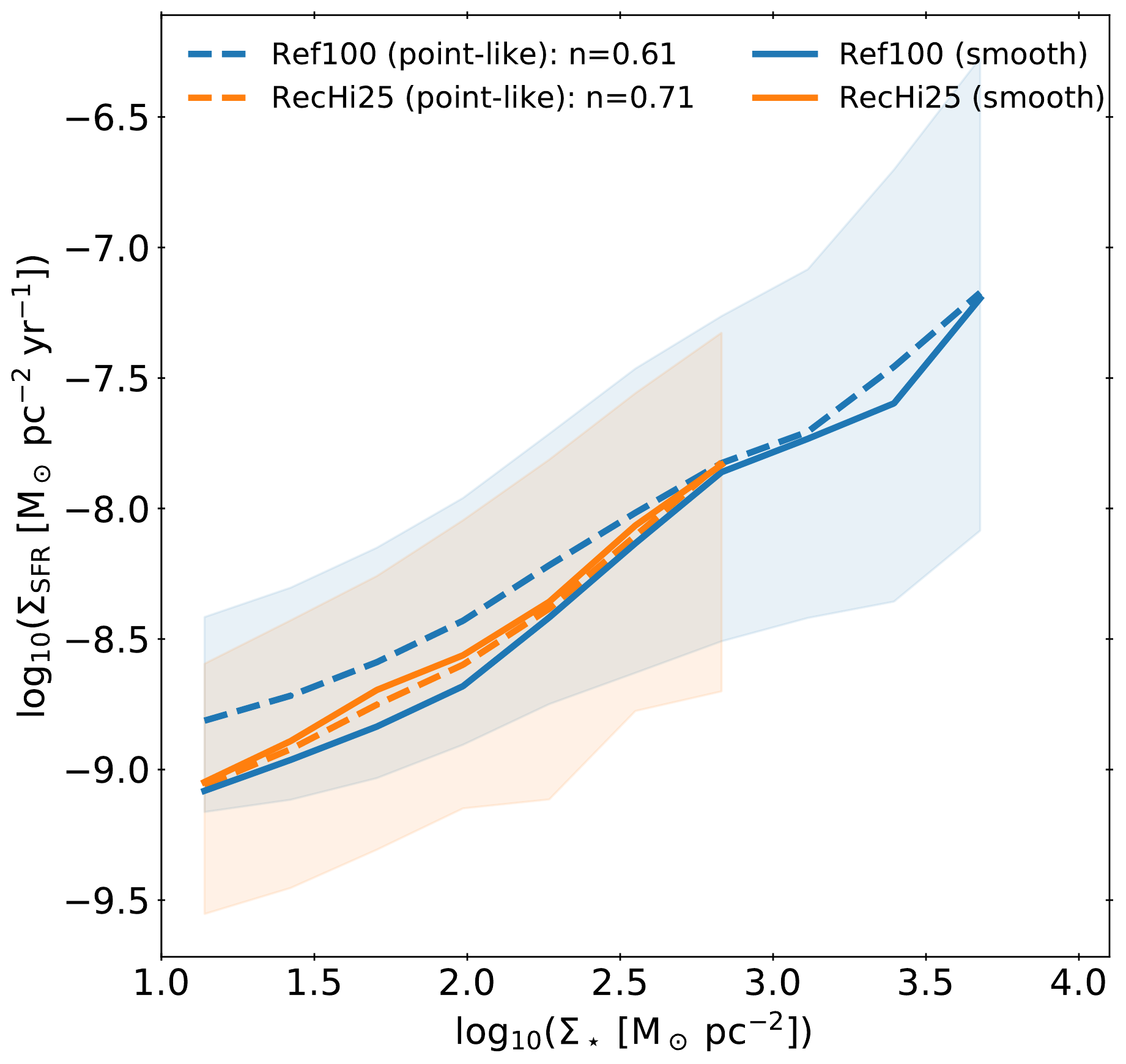}
	\includegraphics[width=0.48\textwidth]{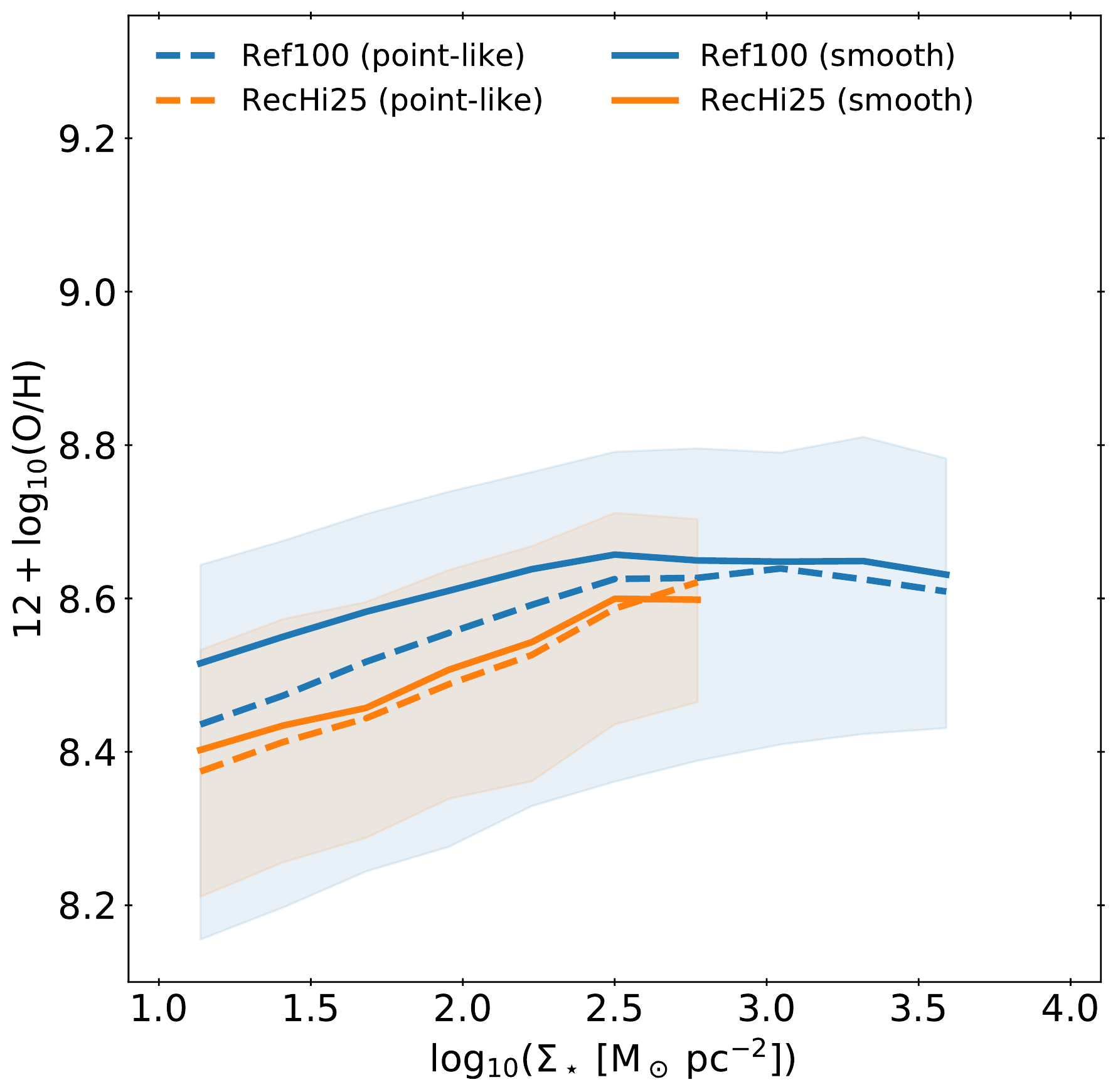}
    \caption{The rSFMS as in Fig.~\ref{fig:sfms} (left panel) and the rMZR relation as in Fig.~\ref{fig:mz} (right panel), but now using property maps that treat the EAGLE star and gas particles as `point-like'. \textit{Shaded regions} denote the 16th-84th percentile range on the relations produced for a \textit{point-like} particle representation. We find that the intrinsic smoothing has little influence on both relations.}
    \label{fig:pointlike}
\end{figure*}

In Fig.~\ref{fig:pointlike}, we plot the rSFMS and rMZR relations of Figs.~\ref{fig:sfms} and \ref{fig:mz} respectively. Encouragingly, we find that for both scaling relations the intrinsic smoothing has only a small effect. The reason for this is that for the selection criteria we employ (sections~\ref{sec:galsel} and \ref{sec:spaxsel}), our choice of nearest-neighbour particle smoothing is small enough to not effect the spaxel values on $\sim$1~kpc scales. In this sense, the scaling relations are well resolved.

\section{Spaxel passive fractions}
\label{sec:pfrac}

\begin{figure}
	\includegraphics[width=\columnwidth]{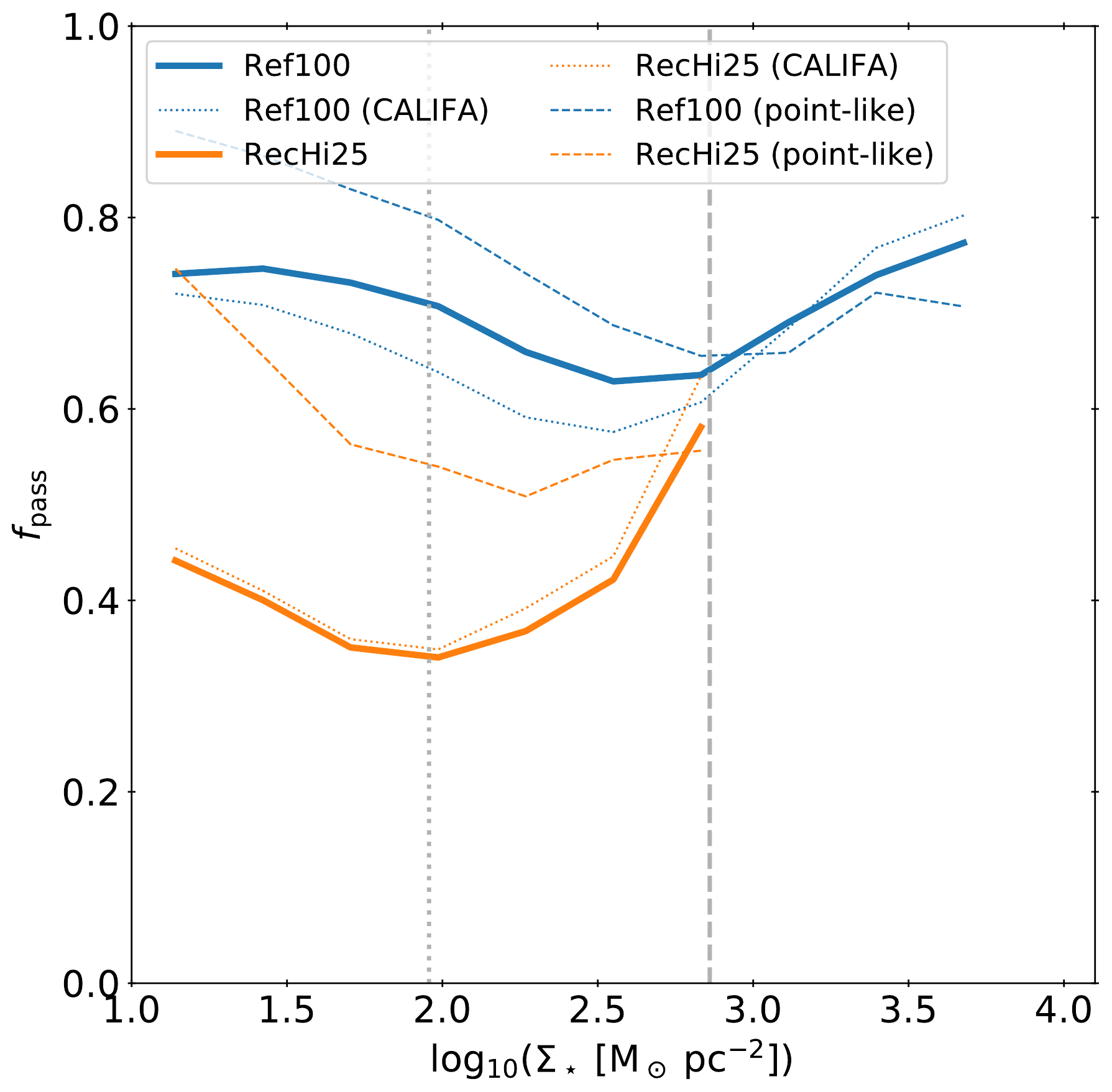}
    \caption{The fractions of spaxels that are passive as a function of $\Sigma_\star$. We plot the weighted (see \S\ref{sec:spaxsel}) fraction for Ref100 (blue) and RecHi25 (green) galaxies. The vertical dashed and dotted lines represent the 10 star-forming particle resolution limit for Ref100 and RecHi25 respectively. We see that the fractions are minimal at their respective resolution limits. The increasing trend in $f_{\rm pass}$ with $\Sigma_\star$ above the resolution limit is taken to be a physical effect, whereas the decreasing trend below the threshold is spurious.}
    \label{fig:pfrac}
\end{figure}

\setcounter{figure}{0}
\makeatletter 
\renewcommand{\thefigure}{D\@arabic\c@figure}
\makeatother
\begin{figure*}	\includegraphics[width=\columnwidth]{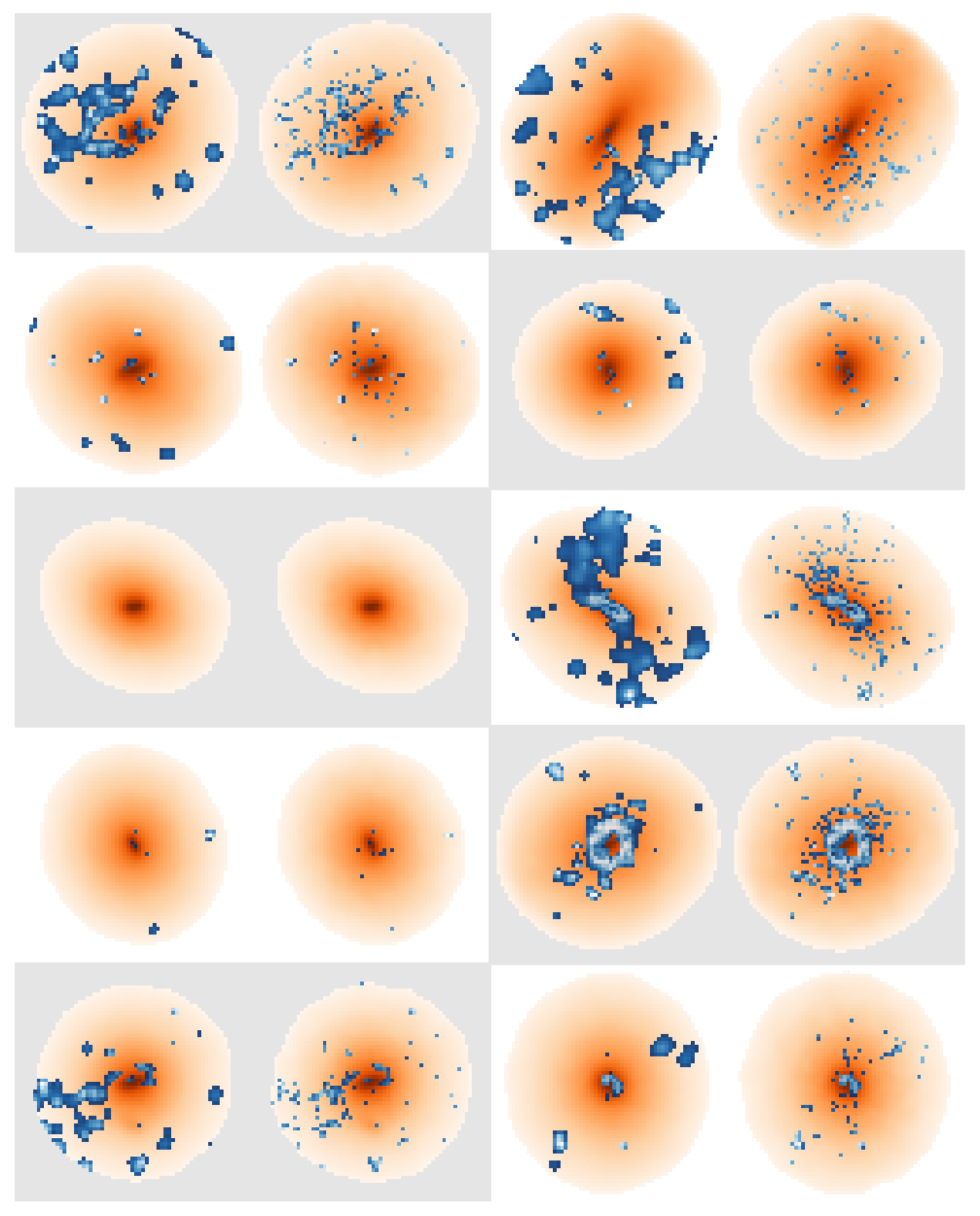}   \includegraphics[width=\columnwidth]{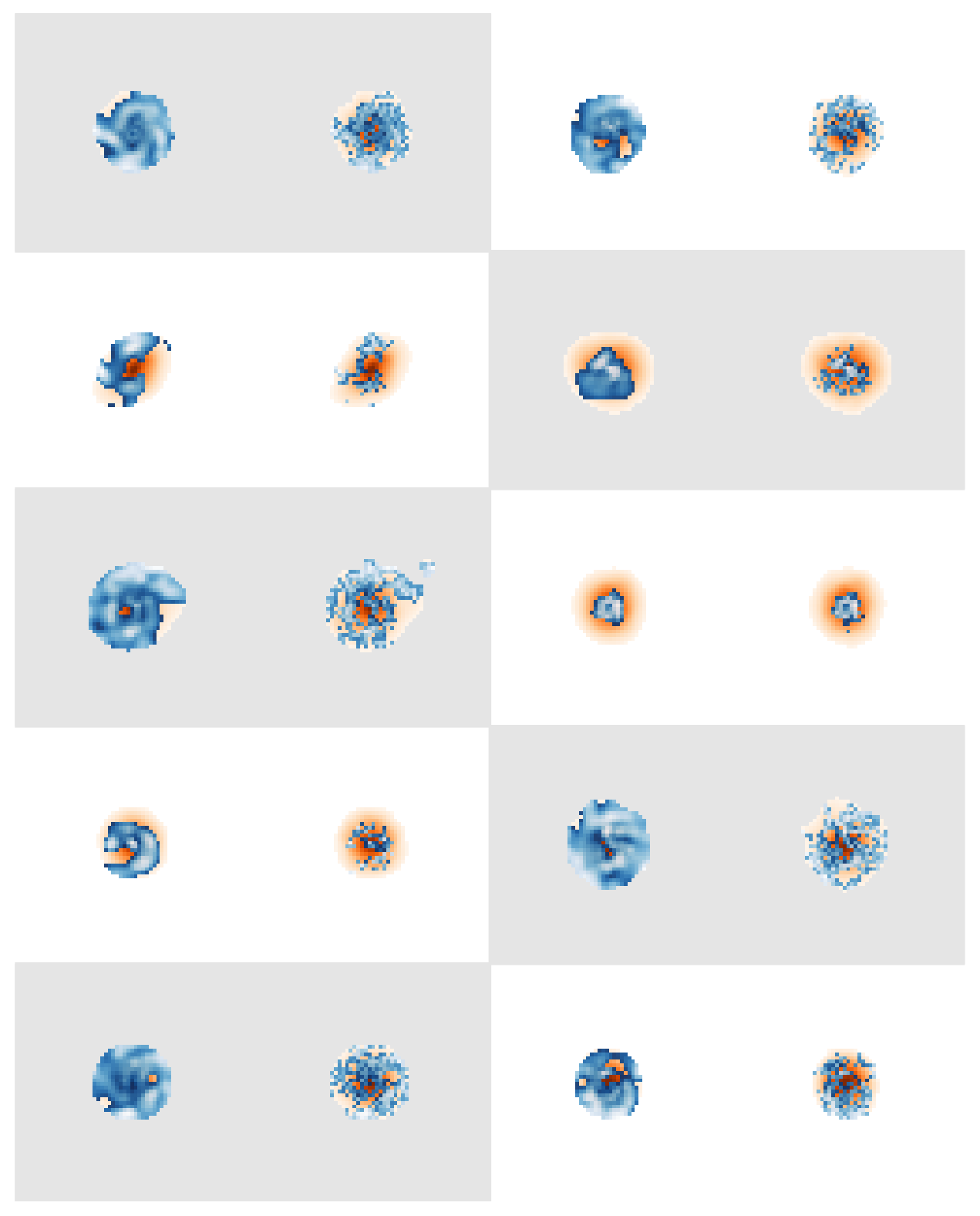}
    \caption{Gallery illustrating the distributions of star-forming gas and stars in Ref100 galaxies. \textit{Orange colour scale} shows \textit{spaxels} shaded from light to dark with increasing $\log_{10} \Sigma_\star$ above 1~${\rm M_\odot \; pc^{-2}}$. \textit{Blue colour scale} shows spaxels shaded from dark to light with increasing $\log_{10}({\rm sSFR}/{\rm yr^{-1}})$ above 10$^{-11}$~${\rm yr}^{-1}$. Pairs of images show the resolved properties made using our standard smoothed maps (left) and from maps where particles are treated as point-like (right). The \textit{left hand panel} shows pairs of images for the 21 most massive EAGLE galaxies, while the \textit{right hand panel} shows 21 galaxies just below the median stellar mass. Comparing the star-forming gas morphologies of the massive and median-mass galaxies gives some insight into the $M_\star$ dependence of the rSFMS seen in Fig.~\ref{fig:splitmass}, revealing a much clumpier gas distribution at high $M_\star$.}
    \label{fig:sfmorph}
\end{figure*}

While the rSFMS and rMZR of star-forming spaxels can elucidate the nature of regions of star formation in galaxies, considering the fraction of passive spaxels provides further insight into how widespread these regions are, and their distribution within the galaxies. The difficulty with computing spaxel \textit{passive fractions}, $f_{\rm pass}$, is that they can be biased by the numerical effects of sampling gas particles in a simulation with finite mass resolution. At low surface densities, this originates from shot noise in the gas particle counts. By considering only the regime where the threshold for passivity is resolved, i.e. corresponding to a sufficient number of gas particles, these numerical effects can be mitigated. 

The relationship between gas particle counts and surface density can be expressed as  

\begin{equation}
\Sigma_{\rm gas} = \frac{N_{\rm parts} m_{\rm gas}}{l_{\rm pix}^2},
\end{equation}

where $N_{\rm parts}$ is the number of particles, $m_{\rm gas}$ is the typical gas particle mass and $l_{\rm pix}$ is the physical spaxel width. \\

To classify spaxels as active and passive, we  must define some threshold level of star formation. We define a spaxel to be active if it is above a local specific SFR of $10 ^ {-11} \, {\rm yr^{-1}}$, matching the $z=0.1$ threshold value used by \citet{Furlong15}. For a given SFR surface density, this corresponds to a stellar mass surface density threshold of:
\begin{equation}
\Sigma_{\rm \star, t} = \frac{\Sigma_{\rm SFR}}{10^{-11} \, {\rm yr^{-1}}}.
\end{equation}
Furthermore, the SFR surface density is closely related to the gas surface density, which is expected due the pressure dependent formulation of the Kennicutt-Schmidt star formation law in EAGLE. We interpolate the median relation, $\tilde{\Sigma}_{\rm SFR}(\Sigma_{\rm gas})$, to obtain the $\Sigma_{\rm \star, t}$ value corresponding to a given gas particle count:
\begin{equation}
\Sigma_\star = 
\tilde{\Sigma}_{\rm SFR}\left(\frac{N_{\rm parts} m_{\rm gas}}{l_{\rm pix}^2}\right) \, 10^{11} \, {\rm yr}.
\end{equation}
For a minimum tolerance of 10 particles at standard resolution, this corresponds to a threshold of $\log_{10} \Sigma_{\rm \star} / ({\rm M_\odot \, pc^{-2}}) > 2.86$. At high resolution this limit is a factor of 8 lower ($\log_{10} \Sigma_{\rm \star} / ({\rm M_\odot \, pc^{-2}}) > 1.96$). 

In Fig.~\ref{fig:pfrac} we plot the passive fraction, $f_{\rm pass}$, as a function of $\Sigma_\star$. We plot the threshold $\Sigma_\star$ values as vertical dotted and dashed lines, corresponding to high and standard resolution, respectively. We see that despite the monotonic increase with $\Sigma_\star$ observed in the rSFMS (Fig.~\ref{fig:sfms}), the passive fraction shows a minimum at the $\Sigma_\star$ that corresponds approximately to the 10 particle resolution limit. The trend of increasing $f_{\rm pass}$ with  $\Sigma_\star$ above the threshold is thus deemed to be a physical effect, demonstrating the increasing passivity in the dense central regions of galaxies. Conversely, the trend of decreasing $f_{\rm pass}$ with  $\Sigma_\star$ trend  below the threshold is spurious, and driven by the poor numerical sampling of star-forming gas particles. 

Combining the portions of the Ref100 and Recal25 trends that are above the resolution limit, we see that the passive fraction increases with $\Sigma_\star$ for $\Sigma_\star \gtrsim 10^{2} \; {\rm M_\odot} {\rm pc}^{-2}$ (and possibly also at lower $\Sigma_\star$). Taken together, the simulations indicate that the overall passive fraction increases by a factor of $\approx$1.3 over the $2 \lessapprox \log_{10}\Sigma_\star/ ({\rm M_\odot} \; {\rm pc}^{-2}) \lessapprox 3.5$ range.%

The problem with the calculation of $f_{\rm pass}$ and specific SFR comes about due to the consideration of active and passive fractions when calculating the values, unlike the rSFMS and rMZR relations where the average values of only star-forming spaxels are considered. When considering average radial profiles affected by the passive threshold resolution criterion, such as specific SFR and $f_{\rm pass}$ profiles, we can indicate the radii where the average stellar mass surface density is above the threshold value. This approach is employed and discussed in \S\ref{sec:sfmsev}. 

\section{Star-forming gas morphologies}
\label{sec:gasmorph}

In Fig.~\ref{fig:sfmorph} we show the morphology of the star-forming gas component in the 21 most massive galaxies in the Ref100 simulation, and the 21 galaxies immediately below the median mass of the $M_\star > 10^{10} {\rm M_\odot}$ sample. Two images are shown for each galaxy, where particles are either smoothed or treated as point-like. We see that the highest-mass galaxies tend to exhibit a much more disrupted and clumpy star forming gas distribution than the median mass galaxies.

\bsp	%
\label{lastpage}
\end{document}